\documentclass[11pt]{article}
\pdfoutput=1

\usepackage{cancel,slashed}
\usepackage{bbm}
\usepackage{bm}
\usepackage{amsmath,amsfonts,amssymb,mathtools,nicefrac,nccmath,cases}
\usepackage{graphicx}
\usepackage{cite}
\usepackage{skak}

\usepackage{dsfont}
\usepackage{latexsym}
\usepackage{color}
\usepackage[hyperfootnotes=false,linktocpage]{hyperref}
\usepackage{transparent}
\usepackage[hang,flushmargin]{footmisc}
\usepackage{blkarray}
\usepackage{multirow}
\usepackage{empheq}
\usepackage{comment}
\usepackage{wasysym}
\usepackage{tikzsymbols}

\usepackage{caption}
\usepackage{subcaption}
\usepackage[normalem]{ulem}

\newcommand{\bmat}{\left(\begin{array}}
\newcommand{\emat}{\end{array}\right)}

\def\R{\mathbbm{R}}

\def\a {\alpha}
\def\b {\beta}
\def\l {\lambda}
\def\m {\mu}
\def\n {\nu}
\def\r {\rho}
\def\s {\sigma}
\def\t {\tau}
\def\bK {{\bf K}}

\def\1{{\bf 1}}
\def\2{{\bf 2}}
\def\3{{\bf 3}}
\def\4{{\bf 4}}
\def\6{{\bf 6}}

\def\targ#1#2{\genfrac{[}{]}{0pt}{}{#1}{#2}}
\def\targ2#1#2{\genfrac{}{}{0pt}{}{#1}{#2}}

\definecolor{mygr}{rgb}{0,0.6,0}
\definecolor{mygrey}{rgb}{0,0.1,0.2}
\definecolor{myblue}{rgb}{0,0.5,0.9}
\definecolor{myblue2}{rgb}{0,0.5,0.5}
\definecolor{myblue3}{rgb}{0,0.7,0.9}
\definecolor{myblue4}{rgb}{0,0.6,0.6}
\definecolor{myorange}{rgb}{1,0.5,0}
\definecolor{mypurple}{rgb}{0.6,0,1}
\definecolor{mygolden}{rgb}{1,0.8,0.2}
\definecolor{mycyan}{rgb}{0,1,1}
\definecolor{mymagenta}{rgb}{1,0,1}
\definecolor{mykiwi}{rgb}{0.8,1,0.5}
\definecolor{mybrown}{cmyk}{0.14, 0.42, 0.56, 0.2}
\definecolor{myturq}{cmyk}{0.99, 0, 0.2, 0.4}
\definecolor{myaubergine2}{cmyk}{0.4, 0.5, 0, 0.1}
\definecolor{myaubergine}{cmyk}{0.6,0.85,0,0}
\definecolor{CycleGreen}{cmyk}{0.52,0,1,0}
\definecolor{CycleBrown}{cmyk}{0, 0.4, 0.9, 0.2}

\DeclareFontFamily{U}{rcjhbltx}{}
\DeclareFontShape{U}{rcjhbltx}{m}{n}{<->rcjhbltx}{}
\DeclareSymbolFont{hebrewletters}{U}{rcjhbltx}{m}{n}

\DeclareMathSymbol{\lamed}{\mathord}{hebrewletters}{108}
\DeclareMathSymbol{\mem}{\mathord}{hebrewletters}{109}
\DeclareMathSymbol{\ayin}{\mathord}{hebrewletters}{96}
\DeclareMathSymbol{\tsadi}{\mathord}{hebrewletters}{118}
\DeclareMathSymbol{\qof}{\mathord}{hebrewletters}{113}
\DeclareMathSymbol{\resh}{\mathord}{hebrewletters}{114}
\DeclareMathSymbol{\pe}{\mathord}{hebrewletters}{112}
\DeclareMathSymbol{\pesofit}{\mathord}{hebrewletters}{80}
\DeclareMathSymbol{\samekh}{\mathord}{hebrewletters}{115}
\DeclareMathSymbol{\tav}{\mathord}{hebrewletters}{116}
\DeclareMathSymbol{\vav}{\mathord}{hebrewletters}{119}
\DeclareMathSymbol{\het}{\mathord}{hebrewletters}{120}
\DeclareMathSymbol{\yod}{\mathord}{hebrewletters}{121}
\DeclareMathSymbol{\zayin}{\mathord}{hebrewletters}{122}
\DeclareMathSymbol{\alephdot}{\mathord}{hebrewletters}{128}
\DeclareMathSymbol{\tsadisofit}{\mathord}{hebrewletters}{90}
\DeclareMathSymbol{\shin}{\mathord}{hebrewletters}{152}

\def\CN {{\cal N}}

\def\CK {{\cal K}}
\def\CV {{\cal V}}

\def\cV {{\cal V}}
\def\cD {{\cal D}}

\def\del{{\delta}}
\def\d {{\rm d}}
\def\be{\begin{equation}}
\def\ee{\end{equation}}
\def\bea{\begin{eqnarray}}
\def\eea{\end{eqnarray}}
\def\bes{\begin{subequations}}
\def\ees{\end{subequations}}

\def\oh{\frac{1}{2}}

\def\pa {\partial}

\def\om{\omega}

\usepackage{multicol}
\usepackage{float}
\usepackage{caption}
\def\r {{\rho}}

\def\p {{\partial}}

\def\g {{\gamma}}

\def\s {{\sigma}}

\def\CO {{\cal O}}

\newcommand{\cF}{\mathcal{F}}
\newcommand{\cG}{\mathcal{G}}

\newcommand{\cK}{\mathcal{K}}

\newcommand{\cN}{\mathcal{N}}
\newcommand{\cO}{\mathcal{O}}

\newcommand{\cT}{\mathcal{T}}
\newcommand{\cI}{\mathcal{I}}

\newenvironment{eqn*}{\begin{equation*}\begin{aligned}}{\end{aligned}\end{equation*}\noindent}

\makeatletter
\newsavebox\myboxA
\newsavebox\myboxB
\newlength\mylenA

\newcommand*\xoverline[2][0.75]{%
\sbox{\myboxA}{$\m@th#2$}%
\setbox\myboxB\null
\ht\myboxB=\ht\myboxA%
\dp\myboxB=\dp\myboxA%
\wd\myboxB=#1\wd\myboxA
\sbox\myboxB{$\m@th\overline{\copy\myboxB}$}
\setlength\mylenA{\the\wd\myboxA}
\addtolength\mylenA{-\the\wd\myboxB}%
\ifdim\wd\myboxB<\wd\myboxA%
   \rlap{\hskip 0.5\mylenA\usebox\myboxB}{\usebox\myboxA}%
\else
    \hskip -0.5\mylenA\rlap{\usebox\myboxA}{\hskip 0.5\mylenA\usebox\myboxB}%
\fi}
\makeatother

\begin{document}
\pagestyle{plain}

\makeatletter
\@addtoreset{equation}{section}
\makeatother
\renewcommand{\theequation}{\thesection.\arabic{equation}}

\pagestyle{empty}
\rightline{IFT-UAM/CSIC-25-45}
\vspace{0.5cm}
\begin{center}
\Huge{{Curvature divergences in 5d ${\cal N}=1$ supergravity} 
\\[10mm]}
\Large{Alejandro Blanco, Fernando Marchesano and Luca Melotti}\\[12mm]
\small{
Instituto de F\'{\i}sica Te\'orica UAM-CSIC, c/ Nicol\'as Cabrera 13-15, 28049 Madrid, Spain
\\[10mm]} 
\small{\bf Abstract} \\[5mm]
\end{center}
\begin{center}
\begin{minipage}[h]{15.0cm}

We study the scalar curvature $R$ of the vector moduli space of 5d $\cN=1$ supergravities, obtained by compactifying M-theory on a Calabi--Yau three-fold. We find that $R$ can only diverge at points where some gauge interactions go to infinite coupling in Planck units and become SCFTs or LSTs decoupled from gravity and other vector multiplets. For 5d SCFTs of rank $r\leq 2$  divergences occur if, additionally, the SCFT still couples to the vevs of such vector multiplets, so that along its Coulomb branch its gauge kinetic matrix and/or string tensions depend on some non-dynamical parameters. If the strong coupling singularity is better understood as a 6d $(1,0)$ SCFT, as in some decompactification limits, then divergences in $R$ arise when the SCFT is endowed with a non-Abelian gauge group.

\end{minipage}
\end{center}
\newpage
\setcounter{page}{1}
\pagestyle{plain}
\renewcommand{\thefootnote}{\arabic{footnote}}
\setcounter{footnote}{0}


\tableofcontents


\section{Introduction}
\label{s:intro}

In recent years, there has been a renewed interest in moduli space singularities arising from string compatifications. In particular, in the context of the Swampland Programme \cite{Vafa:2005ui,Brennan:2017rbf,Palti:2019pca,vanBeest:2021lhn,Grana:2021zvf,Agmon:2022thq} those singularities that correspond to asymptotic infinite-distance limits of an Effective Field Theory (EFT) deserve special attention. The Swampland Distance Conjecture (SDC) \cite{Ooguri:2006in} posits that an infinite tower of states must become light exponentially fast along a geodesic trajectory of this sort. Understanding the nature  of this tower, its features,  and its implications for the physics of the limit has triggered a wealth of activity in the current string theory literature. 

A fruitful setting in which this analysis has been developed is 4d $\cN=2$ supergravity EFTs, obtained from compactifying type II string theory on a Calabi--Yau (CY) three-fold. In particular, in \cite{Grimm:2018ohb,Grimm:2018cpv,Corvilain:2018lgw,Lee:2019wij} powerful mathematical results were applied to understand the classification of asymptotic moduli space singularities from a physical perspective. In fact, one may extend this analysis to include  finite-distance singularities, like conifold points, which also correspond to loci where the moduli space curvature blows up \cite{Strominger:1995cz}. As pointed out in \cite{trenner2010asymptotic},  some asymptotic limits may also lead to moduli space curvature divergences and,  motivated by this fact, \cite{Marchesano:2023thx, Marchesano:2024tod,Castellano:2024gwi,Castellano:2026bnx} initiated a systematic study of curvature divergences in the vector multiplet moduli space of type II CY compactifications.\footnote{See \cite{Raman:2024fcv,Delgado:2024skw,Grimm:2025lip} for related approaches.} The emerging  picture is that curvature divergences are sourced by EFT subsectors that display $\cN=2$ rigid supersymmetry and that decouple from gravity.  

Indeed, using results from special geometry  one can express the moduli space scalar curvature as a sum of two pieces. The first one is a negative constant piece, while the second one is a positive moduli-dependent piece that can diverge at certain points in moduli space.  Restricted to  regions of weak coupling, there are essentially two different ways in which such a term can blow up \cite{Castellano:2024gwi}. The first one is by inducing poles in the cubic derivative of the prepotential, as it happens at conifold points, where the pole represents a massless hyper. The second one is by considering an infinite-distance limit that contains a rigid field theory (RFT) with non-vanishing curvature, and whose gauge interactions become parametrically stronger than gravity along the limit. This  led to propose the Curvature Criterion \cite{Marchesano:2023thx}, which posits that a curvature divergence  at infinite distance signals the presence of an RFT decoupled from gravity. Moreover, it turns out that the nature of the SDC tower restricts the kind of RFT that we can have, together with its UV description as a rigid theory (UVRT). In the case of genuine equi-dimensional emergent string limits \cite{Lee:2019wij} the RFT corresponds to a 4d Seiberg--Witten theory \cite{Castellano:2026bnx}. For decompactification limits the RFT flows to  a 5d or 6d SCFT in the UV, while for higher-dimensional emergent string limits one can also have  a Little String Theory (LST) as an UVRT \cite{Marchesano:2024tod}. 

In this paper we initiate the study of moduli space curvature divergences in the context of 5d $\cN=1$ supergravity. More precisely, we analyse the  scalar curvature of the vector multiplet moduli space of 5d EFTs that arise from compactifying M-theory on a Calabi--Yau three-fold \cite{Cadavid:1995bk,Ferrara:1996hh,Ferrara:1996wv}, and which have been recently considered in \cite{Corvilain:2018lgw,Lee:2019wij,Heidenreich:2020ptx,Katz:2020ewz,Alim:2021vhs,Cota:2022maf,Gendler:2022ztv,Rudelius:2023odg,Kaufmann:2024gqo} in the context of the Swampland Programme. While this is a concrete realisation of 5d supergravity EFTs, it has been recently argued in \cite{Katz:2020ewz,Kaufmann:2024gqo} that the consistency conditions of BPS strings in 5d $\cN=1$ supergravity allows one to reconstruct the features of the vector multiplet of CY compactifications from a bottom-up perspective. In this sense, we expect our results to be equally general. 

Even if the moduli space metric of an M-theory CY compactification is much simpler than its type II analogue, one lacks the machinery of special geometry to write down a simple expression for the curvature. Still, we manage to derive an expression for the 5d scalar curvature that resembles the 4d case. We again obtain a scalar curvature that is a sum of a negative constant piece plus a moduli-dependent piece, which essentially corresponds to the curvature of a 5d rigid theory. Further analysis shows that this second piece can only blow up at points where a vector multiplet subsector  {\em i)} becomes a 5d RFT and {\em ii)} goes to infinite coupling. These two conditions guarantee that such a subsector --  dubbed {\it core RFT} -- decouples from gravity, in agreement with the Curvature Criterion. It is however important to stress that the existence of a core RFT is just a necessary condition for a curvature divergence to develop, and not a sufficient one. 

To understand which additional conditions are needed for a curvature divergence to occur, we analyse trajectories of finite and infinite distance where a core RFT decouples from gravity. In the case of finite-distance trajectories, the core RFT corresponds to $r$ divisors contracting to a point, where a 5d SCFT of rank $r$ is localised. We obtain analytic expressions for the divergent terms of the scalar curvature  in those cases where $r \leq 2$, and which we then evaluate along the Coulomb branch of the 5d SCFT point.  Away from the SCFT point, both the gauge kinetic matrix of the core RFT and its 5d monopole string tensions become finite, and they may then depend on the scalar vevs of the remaining vector multiplets, which from the 5d SCFT viewpoint are perceived as non-dynamical mass parameters. We say that the 5d SCFT (or core RFT) fully decouples from the rest of the vector multiplets of the compactification when both the kinetic matrix and the string tensions are independent of any mass parameters. In that case, there is no curvature singularity at the SCFT point. In contrast, if any of these quantities depends on the mass parameters, then the scalar curvature diverges as we approach the 5d SCFT point. Remarkably, the degree of divergence is sensitive to the kind of dependence that the 5d SCFT can have. If the gauge kinetic matrix physically depends on the mass parameters we obtain a maximal curvature divergence along the Coulomb branch trajectory. If there is a choice of mass parameters where such dependence disappears, but the 5d monopole string tension still depends on them, then we still have a curvature divergence, although milder than the previous one.  

The analysis of infinite-distance limits is a bit more involved, in the sense that it depends on the interplay of the core RFT with the SDC tower. In the case of 6d decompactification limits, the Calabi--Yau must be elliptically fibered, and the interplay translates in the kind of divisors (in terms of the fibration) hosting the core RFT. If the core RFT corresponds to a set of vertical divisors, then there is no divergence, since in this case the UVRT consists of a 6d SCFT with only a tensor branch, and as a consequence the core RFT has a flat metric. If instead the core RFT contains some exceptional divisors, then the UVRT corresponds to a 6d SCFT endowed with a non-Abelian gauge group, and divergences appear. For the case of fibral divisors we have that RFT monopole string tensions lie below the 6d KK scale. As a result, the UVRT of the core RFT corresponds to a 5d SCFT, despite the decompactification limit, and as such one should apply the same criteria as for finite-distance trajectories. Finally, for emergent string limits curvature divergences appear iff the core RFT has a diverging curvature. It follows that for low-rank core RFTs and certain class of limits the divergences are absent. Since in this case the UVRT oftentimes corresponds to an LST, we expect that this statement can be translated into well-defined features of LSTs.

In summary, similarly to previous analysis carried out for 4d $\CN=2$ EFTs, we find that the behaviour of the scalar curvature at boundaries of the moduli space of 4d $\cN=1$ supergravities contains valuable information, both at the IR level, regarding EFT subsectors that decouple from gravity, and at the level of their UV rigid completion. Pursuing the analysis initiated in this work may sharpen this picture, and give us the final hint on which UV information is encoded in the moduli space curvature of those EFTs that are consistent with quantum gravity.

The rest of the paper is organised as follows. In section \ref{s:Mtheory} we describe the vector multiplet sector of M-theory compactified on a CY and compute the scalar curvature of its moduli space. In section \ref{s:rigid} we describe the necessary conditions for curvature divergences in this sector, which involves defining rigid limits in terms of charge-to-tension ratios of 5d BPS strings, and then core RFTs. We obtain explicit expressions for the divergent curvature terms for core RFTs of rank $r \leq 2$, which we then apply to finite-distance trajectories in section \ref{s:w3} and to infinite distance ones in section \ref{s:w21}. In each case, we obtain the necessary conditions for a curvature divergence, and their meaning in terms of UVRT physics. Section \ref{s:examples} illustrates our general findings via a couple of concrete CY examples, and in section \ref{s:conclu} we draw our conclusions. Finally, some technical details are relegated to the appendices. Appendix \ref{ap:asymptotic} describes the asymptotic expansion of the supergravity gauge kinetic matrix and its rigid counterpart, in order to show that the presence of a core RFT is a necessary element for a curvature divergence. Appendix \ref{ap:metric} shows that, with our definition of rigid limit, the moduli space metric reduces to that of a rigid 5d theory.


\section{M-theory on Calabi--Yau three-folds}
\label{s:Mtheory}

In this section we describe the vector multiplet sector of a 5d $\cN=1$ supergravity EFT obtained from compactifying M-theory on a Calabi--Yau three-fold, following the same conventions as in \cite{Alim:2021vhs,Cota:2022maf}. We then derive an explicit expression for the scalar curvature of its moduli space, in order to analyse its divergences in the next section.

\subsection{Vector multiplet moduli space}
\label{ss:moduli}

Let us consider M-theory compactified on a Calabi--Yau  three-fold $X$. We assume an 11d metric of the form
\be
ds^2_{11} = ds^2_{\mathbb{R}^{1,4}} + ds^2_X \, ,
\ee
where $ds^2_X$ is the internal CY metric, encoded in a K\"ahler 2-form $J$ and a holomorphic 3-form $\Omega$. Taking a basis of integral harmonic 2-forms $\ell_{11}^{-2} \omega_a, a=1,...,n=h^{(1,1)}(X)$, with $\ell_{11}$ the 11d Planck length, we can expand the K\"ahler form as
\be
J = M^a \omega_a \, .
\ee
Here $M^a$ are the K\"ahler moduli of $X$, which control the volumes of holomorphic curves and divisors of $X$ in $\ell_{11}$ units. From the viewpoint of the 5d EFT, they describe the scalar fields of the vector multiplet sector, except for the overall volume of $X$, which belongs to the hypermultiplet sector. In terms of the triple intersection numbers of $X$, defined as
\be
\cK_{abc} = \ell_{11}^{-6} \int_X \omega_a \wedge \omega_b \wedge \omega_c \, ,
\ee
the total Calabi--Yau volume in 11d Planck units reads
\be
\cV_X = \frac{1}{6} \cK_{abc} M^a M^b M^c \, .
\ee
To describe the vector multiplet moduli space, it is customary to introduce a set of coordinates $\psi^i$, $i=1,...,n-1$ that parametrise the constant-volume hypersurface  $\cV_X =1$. In addition to them, the vector  multiplet sector of the 5d theory contains $U(1)$ gauge fields $A^a$, with field strengths $F^a = dA^a$, that arise from expanding the 11d 3-form in the basis of harmonic 2-forms
\be
C_3 = A^a \wedge \omega_a \, .
\ee
The bosonic 5d action for this sector then can be written as \cite{Lauria:2020rhc}
\be     \label{SVM5d2}
\begin{split}
S_{\rm 5d}^{\rm VM} & =  \frac{2\pi}{\ell_5^3} \int_{\mathbb{R}^{1,4}} \left(R_5 * \mathbbm{1} - \mathfrak{g}_{ij} d\psi^i \wedge * d\psi^j\right)  - \frac{1}{4\pi \ell_5}   \int_{\mathbb{R}^{1,4}} {\cal I}_{ab} F^a \wedge *_5 F^b \\ & -  \frac{1}{24\pi^2}  \int_{\mathbb{R}^{1,4}} \cK_{abc}\, A^a \wedge F^b \wedge F^c\, ,
\end{split}
\ee
where $\ell_5$ is the five-dimensional Planck length (related to the 5d Planck mass by $M_{\rm P}^3=\frac{4\pi}{\ell_5^3}$). The triple intersection numbers $\cK_{abc}$ describe the Chern-Simons (CS) couplings, while the gauge kinetic matrix ${\cal I}_{ab}$ is given by 
\be
{\cal I}_{ab} = - \,\pa_{M^a} \pa_{M^b} \log \cV_X \Big|_{\cV_X=1} = \frac{1}{4} \cK_a \cK_b - \cK_{ab} \, ,
\label{Iabsugra}
\ee
where we have defined $\cK_a \equiv \cK_{abc}M^b M^c$ and $\cK_{ab} \equiv \cK_{abc}M^c$. Finally, the moduli space metric is nothing but the induced metric on the constant-volume hypersurface
\be
\mathfrak{g}_{ij} = \frac{1}{2} \cI_{ab}\, \pa_{i} M^a \pa_{j} M^b\, ,
\label{metric}
\ee
and its inverse is related to the gauge kinetic matrix as
\be
\cI^{ab} = \frac{1}{2} \mathfrak{g}^{ij}\pa_{i} M^a \pa_{j} M^b + \frac{1}{3} M^aM^b\, .
\label{invrel}
\ee

If we further compactify the theory on a circle, we obtain a 4d ${\cal N}=2$ theory that corresponds to type IIA string theory on $X$. Here the vector multiplet moduli are complexified
\be
T^a = b^a + i t^a \, ,
\ee
by adding an axionic part $b^a$ that comes from reducing the 5d gauge fields $A^a$ on the circle, while the saxions $t^a$ control the volume of the 2-cycles of $X$ in string units and are related to the M-theory moduli as
\be
M^a = \frac{t^a}{\cV_{\rm IIA}^{1/3}}\, .
\label{Mt}
\ee
Note that in the type IIA setup the overall volume $\cV_{\rm IIA} = \frac{1}{6} \cK_{abc} t^a t^b t^c$ also belongs to the vector multiplet sector and the constant-volume constraint does not need to be imposed. The 4d bosonic action reads
\be
S_{\rm 4d}^{\rm VM} =  \frac{1}{2\kappa_{4}^2} \int_{\R^{1,3}} R_4 * \mathbbm{1} - \cG_{ab} \left( db^a \wedge * db^b + dt^a \wedge * dt^b \right) + \text{gauge fields} \, ,
\label{SVM}
\ee
In the large volume approximation, the 
 moduli space metric only depends on the saxions and can be written as
\be
\cG_{ab} = -\frac{1}{2} \pa_{t^a} \pa_{t^b} \log \cV_{\rm IIA}  = \frac{1}{2} \cV_{\rm IIA}^{-2/3} {\cI_{ab}}  \, .
\label{IIAmetric}
\ee
This relation between 5d and 4d couplings turns out to be quite useful in order to compute the 5d moduli space curvature, as we now turn to discuss.

\subsection{The scalar curvature}
\label{ss:curvature}

In order to compute the scalar curvature of the 5d vector multiplet scalar metric \eqref{metric} we will consider an auxiliary metric with a similar one. Such a metric is nothing but the type IIA large-volume metric that appears in \eqref{SVM}, restricted to the real $n$-dimensional saxionic slice of the moduli space. We dub the corresponding curvature as IIA saxionic curvature. 

To show that the IIA saxionic scalar curvature coincides with the 5d supergravity curvature, let us start with the metric $\cG_{ab}$ in \eqref{IIAmetric} restricted to the saxionic coordinates $t^a$, and perform a change of coordinates to $\chi^I = (\chi^0, \chi^i)$, where
\be
\begin{split}
\chi^0 & = \frac{1}{\sqrt{6}} \log \cV_{\rm IIA} \, , \\
\chi^i & = \chi^i(t^a)\, ,
\end{split}
\label{chicoord}
\ee
with $\chi^i$ homogeneous functions of zero degree of the type IIA saxions $t^a$, such that $\det \frac{\pa \chi}{\pa t} \neq 0$. In these new coordinates the saxionic metric reads
\be
\tilde{\cG}_{IJ} =
\begin{pmatrix}
    1 & 0 \\
    0 & \tilde{\cG}_{ij}
\end{pmatrix}\, 
\ee
where $\tilde{\cG}_{ij} = \frac{\pa t^a}{\pa \chi^i} \frac{\pa t^b}{\pa \chi^j} \cG_{ab}$ is the induced metric on the constant volume hypersurface. In other words, the overall volume decouples from the rest of the coordinates. This can be expected on physical grounds, because in type IIA vector multiplet moduli space $\chi^0$ can be identified with the 10d dilaton, but also directly by computing the following inverse metric elements 
\be
\tilde{\cG}^{00} = \frac{1}{4 \cV_{\rm IIA}^2} \cK^{\rm IIA}_c \cK^{\rm IIA}_d \cG^{cd} = 1 \,, \qquad \tilde{\cG}^{0i} \propto \cK^{\rm IIA}_c \cG^{cd} \frac{\pa \chi^i}{\pa t^d} \propto t^d \frac{\pa \chi^i}{\pa t^d} =0\, ,
\ee
where $\cK_a^{\rm IIA} \equiv \cK_{abc}t^b t^c$ and we have used the zero-degree homogeneity of $\chi^i$. In addition, it is easy to see that the metric $\tilde{\cG}_{ij}$ is invariant under a homogeneous rescaling of the saxions $t^a \to \l t^a$, which means that it only depends on the coordinates $\chi^i$ and not on the overall volume. This implies that the scalar curvature of $\cG_{ab}$ is the same as that of $\tilde{\cG}_{ij}$, which by construction is a zero-degree homogeneous function of the saxionic coordinates $t^a$.

Finally, because the coordinates $\chi^i$ are of degree zero on the type IIA saxions $t^a$, one can identify them with the 5d coordinates $\psi^i$ via $\chi^i(t^a) =\psi^i(t^a/\CV_{\rm IIA}^{1/3})$. Upon this identification we have that $\tilde{\cG}_{ij} = \mathfrak{g}_{ij}$, as can be verified by direct comparison:
\be
\tilde{\cG}_{ij} = \frac{\pa t^a}{\pa \chi^i} \frac{\pa t^b}{\pa \chi^j} \frac{1}{2\cV_{\rm IIA}^{2/3}} \cI_{ab} = \frac{\pa M^a}{\pa \psi^i} \frac{\pa M^b}{\pa \psi^j} \frac{1}{2} \cI_{ab} = \mathfrak{g}_{ij} \, ,
\ee
where we have used \eqref{Mt}, \eqref{IIAmetric} and the fact that  $\cV_{\rm IIA}$ is independent of $\chi^i$. This implies that the curvature of the saxionic slice of the type IIA vector moduli space is equal to the M-theory vector moduli space curvature, associated to the metric $\mathfrak{g}_{ij}$. 

The advantage of this identification is that the IIA saxionic scalar curvature  is easier to compute explicitly. Indeed, since $\cG_{ab}$ can be written as the second derivative of a K\"ahler potential, it satisfies the following property 
\be
\pa_a \cG_{bc} = \pa_b \cG_{ac}\, .
\ee
This in turn allows us to write the IIA saxionic scalar curvature in 4d Planck units explicitly  as 
\be
R_{\rm IIA, sax} = \frac{1}{2} \cG^{ab} \cG^{cd} \cG^{ef} \pa_a \cG_{c[e} \pa_{b]} \cG_{df}\, .
\ee
One can then use the explicit expression for the saxionic metric and its inverse:
\be
\cG_{ab} = \frac{1}{2\cV_{\rm IIA}} \left(\frac{1}{4 \cV_{\rm IIA}} \cK_a^{\rm IIA} \cK_b^{\rm IIA} - \cK_{ab}^{\rm IIA} \right)\, ,
\qquad
\cG^{ab} = t^a t^b - 2\cV_{\rm IIA} \cK_{\rm IIA}^{ab}\, ,
\ee
where we have defined $\cK_{ab}^{\rm IIA} \equiv \cK_{abc}t^c$ and $\cK_{\rm IIA}^{ab}$ as its inverse. Using in addition that
\be\nonumber
\pa_a \cG_{bc} = -\frac{1}{2\cV_{\rm IIA}} \left[ \cK_{abc} -\frac{1}{2\cV_{\rm IIA}} \left( \cK_a^{\rm IIA} \cK_{bc}^{\rm IIA} + \cK_{b}^{\rm IIA} \cK_{ac}^{\rm IIA} + \cK_{c}^{\rm IIA} \cK_{ab}^{\rm IIA} \right) + \frac{1}{4\cV_{\rm IIA}^2} \cK_a^{\rm IIA} \cK_b^{\rm IIA} \cK_c^{\rm IIA} \right] \, ,
\ee
one finally obtains the expression
\be
R_{\rm IIA, sax} = - \frac{1}{2}n \left(n-1\right) + \frac{1}{8\cV_{\rm IIA}^2} \cG^{ab}\cG^{cd}\cG^{ef} \cK_{ac[e} \cK_{b]df}  \, .
\ee

Notice that, as expected, the scalar curvature is a zero-degree homogeneous function of the saxionic coordinates. Expressing it in terms of the variables $M^a$ we obtain the scalar curvature of the M-theory vector moduli space, now in 5d Planck units:
\be     \label{RMth1}
R_{\rm M} = - \frac{1}{2}n \left(n-1\right) + \cI^{ab}\cI^{cd}\cI^{ef} \cK_{ac[e} \cK_{b]df} \, ,
\ee
where $\cI^{ab} = \frac{1}{2} M^aM^b - \cK^{ab}$, with $\cK^{ab}$ the inverse of $\cK_{ab} = \cK_{abc}M^c$. One can compare this formula to the one obtained in section 3 of \cite{Marchesano:2023thx}. Besides the different prefactors and the antisymmetrisation on the triple intersection numbers, the M-theory curvature presents the same structure as its type IIA counterpart. A negative, constant term that grows quadratically with the number of vector multiplets, together with a moduli-dependent term that can diverge whenever some gauge couplings do. This comparison also shows that the field-space curvature divergences present in type IIA large-volume trajectories are comparable to those in their M-theory uplift, although in some cases where the antisymmetrisation of the triple intersection numbers results into a suppressed or lack of divergence in the M-theory counterpart.\footnote{Another instance in which type IIA curvature divergences appear larger that their M-theory counterpart is when the type IIA rigid curvature is generated by world-sheet instanton corrections \cite{Marchesano:2023thx}. However, it follows from the results in \cite{Marchesano:2024tod} that in order to lift such trajectories to M-theory all instanton effects must vanish asymptotically, such that no divergence is generated for the type IIA curvature. By consistency, in these cases there should be no curvature divergence in the M-theory setting, an expectation that we will check in the following sections.\label{ft:instantons}}

In any event, this structure for the moduli space scalar curvature  goes along the lines of the Curvature Criterion proposed in \cite{Marchesano:2024tod}, by which a curvature divergence signals the presence of a non-gravitational, interacting sector that decouples from gravity. To further check this link, it will prove useful to rewrite the last expression as
\be   \label{RMth2}
R_{\rm M}  = - \frac{3}{4}\left(n-2\right)\left(n-1\right) - \cK^{ab} \cK^{cd} \cK^{ef} \cK_{ac[e} \cK_{b]df}
\, .
\ee
 This can be seen as the main result of this section, on which we will base our discussion of the subsequent ones. Based on the results of Appendix \ref{ap:asymptotic}, in the next section we argue that only certain indices can contribute to a divergence in the second term of \eqref{RMth2}. Such indices will give a meaning to the tensor $-\cK^{ab}$ that contracts the triple intersection numbers, namely as the inverse metric (or the squared gauge coupling) of a 5d EFT sector with rigid supersymmetry.


\section{Rigid limits and curvature divergences}
\label{s:rigid}

In this section we study the conditions under which the vector multiplet scalar curvature diverges. For this we apply the philosophy of \cite{Castellano:2024gwi} and define the notion of rigid limit in terms of tension-to-charge ratio of certain 5d BPS strings. We then argue that for a scalar curvature divergence one needs the presence of vector multiplet subsector -- dubbed core RFT -- which goes towards a rigid and strong coupling limit along a given trajectory, which may be of finite or infinite distance. We derive this result in the context of EFT string limits and growth sectors based on them, where one can easily organise the asymptotic curvature in a power expansion with different degrees of divergence. We obtain explicit expressions for the two leading terms of this expansion for the case of core RFTs of rank $r \leq 2$, that will be analysed in the next two sections.

\subsection{EFT string limits and growth sectors}
\label{ss:EFT}

As mentioned above, the M-theory moduli space scalar curvature \eqref{RMth1} shares some properties with the one computed for type IIA string theory compactified on the same Calabi--Yau $X$, in the large volume regime. In particular, the M-theory curvature can only diverge if the saxionic type IIA curvature of the classical metric \eqref{IIAmetric} also does, along the corresponding trajectory in type IIA vector multiplet moduli space. For this reason, to build a representative set of trajectories with curvature divergences, we will study the same class of large-volume vector multiplet limits that were considered in \cite{Marchesano:2023thx,Marchesano:2024tod} in the context of type IIA compactifications, uplifted to a trajectory in M-theory moduli space. To construct them, one first considers a basis of Nef divisor classes $[\ell_s^{-2} \om_a']$ to expand the K\"ahler form as $J = t^a \om_a'$, and then the following set of trajectories for the saxionic type IIA variables $t^a$ in such a basis
\be
t^a = t_0^a + e^a_0 \phi  + \sum_{i=1}^{n-1} e^a_i \phi^{\g_i}  \, , \qquad \phi \to \infty\, .
\label{growth}
\ee
Here $e^a_I \in \mathbb{N}$, $I = (0, i)$ correspond to a set of Nef divisors ${\cal D}_{{\bm{e}}_I} = \ell_s^{-2} e^a_I [\om_a'] = \ell_{11}^{-2} e^a_I [\om_a]$, with the property $\sum_a e^a_I e^a_J = 0$ for $I \neq J$, while $1 > \gamma_1 > \gamma_2 > \dots \geq 0$. Whenever $\gamma_i =0, \forall i$ we say that this limit is an EFT string limit, following the nomenclature of \cite{Lanza:2020qmt,Lanza:2021udy,Lanza:2022zyg}, while we  refer to the general case as a growth sector \cite{Grimm:2018cpv}. The advantage of EFT string limits is that there is a simple dictionary between the topological properties of the divisor class ${\cal D}_{{\bm{e}}_0}$ and the physics of the limit. Indeed, following \cite{Marchesano:2022axe} one can adapt the classification of infinite-distance trajectories made in \cite{Corvilain:2018lgw,Lee:2019wij} to EFT string limits, in order to define the scaling weight $w$ as follows:

\begin{enumerate}

    \item $w=3$ corresponds to the limits  where ${\bf k} \equiv \CK_{abc}e^a_0 e^b_0 e^c_0\neq 0$.
    
    \item $w=2$ limits occur when ${\bf k} = 0$ and ${\bf k}_a \equiv \CK_{abc} e^b_0 e^c_0 \neq 0$ for some $a$. 
 
    \item $w=1$ limits happen whenever ${\bf k}_a  = 0$, $\forall a$. 
    
\end{enumerate}
In the type IIA setup and for EFT string limits, we have that the CY volume grows asymptotically as ${\cal V}_{\rm IIA} \sim \phi^w$ along the limit. Moreover, $w=3$ and $w=2$ limits respectively correspond to decompactification limits to M- and F-theory on $X$, while $w=1$ to emergent string limits \cite{Marchesano:2022axe}. One may also use the same definition for $w$ to classify growth sectors where some $\gamma_i \neq 0$, as long as the leading terms in $\cV_{\rm IIA}$ are still selected by the direction $t^a \propto e_0^a$. While this is the convention that we will apply in the following, one should note that some of the above statements are modified for general growth sectors. In particular ${\cal V}_{\rm IIA}$ may no longer grow as $\phi^w$ for $w=2$ and $w=1$, and some $w=1$ may be nested limits that lead to emergent strings in higher dimensions \cite{Marchesano:2024tod}.

In the following, we will test the asymptotic behaviour of the M-theory scalar curvature \eqref{RMth1} along \eqref{growth}, translated into 5d vector multiplet trajectories via the map \eqref{Mt}. Again, each value of the scaling weight $w$  corresponds to a different class of limits in geometric and physical terms. Limits with weight $w=3$ correspond to trajectories of finite length in 5d vector moduli space, which for the case of EFT string limits take the form
\be
M^a(\phi) = \left(\frac{6}{{\bf k}}\right)^{1/3}  \left( e_0^a + \phi^{-1} \left( t_0^a - e_0^a \frac{{\bf k}_b t_0^b}{{\bf k}}\right)  + \CO(\phi^{-2}) \right) .
\label{Mtrajw3}
\ee
Even if these are finite-distance trajectories, they can give rise to curvature singularities. Following the analysis in \cite{Marchesano:2023thx}, this may only happen when the endpoint of \eqref{Mtrajw3} lies at a finite-distance boundary of the vector multiplet moduli space where $\cI^{ab}$ diverges. At such a boundary, a divisor collapses to a point, signaling the location of a strong coupling singularity \cite{Witten:1996qb}. 
 
Limits with weight $w=2$ and $w=1$ correspond to infinite-distance limits that are either decompactification limits, emergent string limits, or a nested combination of them. Growth sectors with $w=2$ can correspond geometrically to either J-Class A or J-Class B limits in the classification made in \cite{Lee:2019wij}, and always lead to decompactification limits to 6d. Growth sectors with $w=1$ are instead mapped to J-Class B limits in \cite{Lee:2019wij}, and either lead to 5d or 6d emergent string limits. While any infinite distance limit corresponds to a weak coupling limit where some of the entries of $\cI^{ab}$ vanish \cite{Heidenreich:2020ptx}, it may be that simultaneoulsy some other entries blow up, signaling again that we are approaching a strong coupling regime for a subsector of the 5d EFT. Geometrically, one again expects that for this to happen a divisor must collapse to zero size, with different kinds of effective divisors corresponding to different physics \cite{Marchesano:2024tod}. As we  now discuss, a more accurate necessary condition for a curvature divergence is that a subsector of the 5d EFT approaches a strong coupling regime together with a rigid limit, as formulated in \cite{Castellano:2024gwi}.

\subsection{BPS strings and rigid limits}
\label{ss:gamma}

As pointed out in \cite{Castellano:2024gwi} in the context of 4d $\cN =2$ supergravity, the decoupling from gravity of a subsector of an EFT can be detected by looking at the spectrum of charged $\oh$BPS objects. More precisely, when some of the charge-to-mass ratios of such objects diverge along a given limit, one expects to find EFT subsectors whose couplings approach the relations of rigid supersymmetry. Following the nomenclature of  \cite{Castellano:2024gwi} we will dub those limits where a decoupling from gravity occurs as rigid limits, and the decoupling sector as a Rigid Field Theory (RFT). 

In our case, the fundamental $\oh$BPS objects of the  5d EFT correspond to strings and particles. BPS strings magnetically charged under the vector multiplet sector come from wrapping M5-branes on effective divisors ${\cal D}_{\bm{p}} = \ell_{11}^{-2} p^a[\om_a]$ of $X$. Their tension and physical charge are \cite{Alim:2021vhs,Cota:2022maf}
\be
T_{\bm{p}} = 2\pi (4\pi)^{-\frac{2}{3}} \oh p^a \cK_a M_{\rm P}^2 \, , \qquad {\cal Q}_{\bm{p}}^2 = 2\pi (4\pi)^{-\frac{1}{3}} \cI_{ab} p^a p^b M_{\rm P} \, ,
\ee
while their ratio reads \cite{Kaufmann:2024gqo}
\be
\gamma_{\bm{p}}^2 \equiv \frac{{\cal Q}^2_{\bm{p}}}{T_{\bm{p}}^2} M_{\rm P}^3  = \frac{2}{3} + \mathfrak{g}^{ij} \p_i \log (T_{\bm{p}}/M_{\rm P}^2) \p_j \log (T_{\bm{p}}/M_{\rm P}^2)\, ,
\label{gammastring}
\ee
or equivalently
\be
\g_{\bm{p}}^2 = 2  - 8 \frac{\cK_{ab} p^a p^b}{(\cK_c p^c)^2} \, .
\label{gammastring2}
\ee
Notice that the second equality in \eqref{gammastring} comes from \eqref{invrel}, which can in turn be interpreted as a no-force condition for mutually BPS strings
\be
{\cal Q}_{\bm{p}}^2 M_{\rm P}^3 = \frac{2}{3} T_{\bm{p}}^2 + \mathfrak{g}^{ij} \p_i T_{\bm{p}} \p_j T_{\bm{p}}\, ,
\ee
which balances the exchange of gauge bosons, gravitons and moduli, as in 4d settings  \cite{Palti:2017elp,Heidenreich:2019zkl,Lanza:2020qmt}.

Similarly, charged BPS particles come from wrapping M2-branes on effective curves of $X$. If their coupling to the $U(1)$ gauge fields is $q_a A^a$, then their mass and physical charge are given by 
\be
m_{\bm{q}} = 2\pi (4\pi)^{-\frac{1}{3}} q_a M^a M_{\rm P}\, , \qquad \mathfrak{q}^2_{\bm{q}} = 2\pi (4\pi)^\frac{1}{3} \cI^{ab} q_a q_b M_{\rm P}^{-1} \, ,
\ee
while their charge-to-mass ratio reads
\be
\g_{\bm{q}}^2 \equiv M_{\rm P}^3 \frac{\mathfrak{q}^2_{\bm{q}}}{m_{\bm{q}}^2} = \frac{2}{3} + \mathfrak{g}^{ij} \p_i \log (m_{\bm{q}}/M_{\rm P})  \p_j \log (m_{\bm{q}}/M_{\rm P}) 
= 1 - 2 \frac{\cK^{ab}q_a q_b}{(q_c M^c)^2} \, .
\label{gammapart}
\ee
Note that the ratio of a BPS particle with charge $\bm{q}$ equals the ratio of a string with charge $\bm{p}$ if
\be
 q_a \propto \cI_{ab} p^b \, ,
\label{pqdual}
\ee
a relation that, for sufficiently large charges, can be approximated arbitrarily well at any point in moduli space. When  charges at both sides of the relation correspond to effective submanifolds, one may identify the corresponding BPS objects as electro-magnetic duals. Notice in particular that \eqref{pqdual} implies that $m_{\bm{q}}  \propto T_{\bm{p}}/M_{\rm P}$, so along a limit in which a string becomes tensionless there are always candidates for one or several towers of BPS particles that becomes massless, and whose quantised charges are determined by this relation. 

Among the different $\oh$BPS strings present in a compactification of M-theory on a Calabi--Yau $X$, a special role is played by the so-called {\it supergravity} strings \cite{Katz:2020ewz}. These are described as fundamental strings that only appear in gravitational 5d EFTs, whose charge-to-tension ratio $\g_{\bm{p}}$ is expected to be $\cO(1)$ all over the moduli space, and which decouple from the EFT when gravity does. Geometrically, they are identified with M5-branes wrapping Nef divisors of $X$, and they have played a prominent role within the Swampland Programme because consistency of their world-sheet theory implies a set of bounds on the Chern-Simons couplings of the bulk supergravity action \cite{Katz:2020ewz,Kaufmann:2024gqo}.\footnote{A similar story applies to $\oh$BPS objects in other dimensions \cite{Kim:2019vuc,Angelantonj:2020pyr,Martucci:2022krl}. In particular, the 4d version of 5d supergravity strings are the EFT strings featured above, that motivate the trajectories \eqref{growth}.}

In this work we focus our attention on a complementary set of $\oh$BPS fundamental strings, namely those that persist in the spectrum when gravity is decoupled, and that are charged under the decoupling RFT. Along such gravity-decoupling limits their charge-to-tension ratio as defined in \eqref{gammastring} must blow up, since in rigid theories the no-force condition between mutually BPS objects only involves the exchange of gauge bosons and moduli. The exchange of gravitons, controlled by the string tension, should be negligible compared to these two. Hence,  along a rigid limit, a BPS string with magnetic charges $\bm{p}$ only within the RFT must have a tension-to-mass ratio $\gamma_{\bm{p}}$ that diverges asymptotically. From \eqref{gammastring2} we see that this takes us to a regime where
\be
\frac{1}{4} \cK_\mu \cK_\nu \ll -  \cK_{\mu\nu}  \implies \cI_{\mu\nu} \simeq - \cK_{\mu\nu} \, ,
\label{rigid}
\ee
for those field directions $M^\mu$ aligned with the string charge $p^\mu$. This relation for that gauge kinetic matrix is indeed what we expect for a rigid theory, since it amounts to remove the $\log$ from \eqref{Iabsugra} \cite{Bertolini:1994cb,Lauria:2020rhc} in a way that is compatible with charge quantisation.\footnote{Indeed, notice that if we did not care about charge quantisation, we could perform a change of coordinates like \eqref{chicoord} in order to obtain the relation \eqref{rigid} at any point in moduli space.} Simultaneously, the moduli space metric $\mathfrak{g}_{ij}$ reduces to $\mathfrak{g}_{\mu\nu} \simeq - \oh \cK_{\mu\nu}$ for the same coordinates, as shown in Appendix \ref{ap:metric}. For this reason, we dub fundamental strings with diverging $\gamma_{\bm{p}}$ as rigid or RFT strings.

In order to identify rigid limits one may focus in trajectories of the form \eqref{growth} where $\gamma_{\bm{p}} \to \infty$ for some string charge $\bm{p}$.\footnote{Rigid limits will also display $\oh$BPS particles whose charge-to-mass ratio \eqref{gammapart} diverges, as it follows from the relation \eqref{pqdual}. However, just like in 4d $\cN=2$ limits \cite{Castellano:2024gwi}, the dictionary between RFT field directions and electric charges with diverging $\g$'s is much less direct than for magnetic charges, which in this case correspond to strings.} The dictionary between the  string charges whose ratio diverges and the field directions that satisfy \eqref{rigid} is particularly simple when a rigid sector contains the strongest gauge couplings and the  largest divergences in $\gamma_{\bm{p}}$. In that case, the set of string charges $\bm{p}$ with such divergences in $\gamma_{\bm{p}}$ forms a vector subspace, since if $\gamma_{\bm{p}'}$ has a lower divergence so does $\gamma_{\bm{p}+\bm{p}'}$. One can then simply identify the RFT field directions with a basis $\{ \bm{p}_{\rm RFT} \}$ for such a vector subspace, which is equivalent to selecting the subset of vector multiplets under which these rigid strings are magnetically charged.

Interestingly, this case is quite relevant to describe divergences in the moduli space curvature. As already mentioned, \eqref{RMth1} implies that curvature divergences can only be generated at moduli space boundaries with strongly coupled sectors. By the Curvature Criterion \cite{Marchesano:2024tod}, one moreover expects such strongly coupled sectors to decouple from the gravitational sector of the EFT. In particular, one expects that the scalar curvature divergence is sourced by an RFT subsector with diverging gauge couplings, which we dub {\em core RFT}.  This expectation is indeed compatible with  \eqref{RMth2}, which already suggests that the divergence is contained in the rigid piece of the gauge couplings. In the following we discuss in some detail how this picture is realised.

\subsection{RFT sectors and decoupling}
\label{ss:decoupling}

Let us first discuss how the gravitational sector of the EFT, defined as those field directions $M^a \propto p^a$ with  asymptotically finite $\gamma_{\bm{p}}$, decouple from the RFT sectors of a rigid limit. By looking at the  behaviour of the gauge kinetic functions and string charge-to-tension ratios along different limits, as done in Appendix \ref{ap:asymptotic}, one realises that for any trajectory of the form \ref{growth}  -- translated into M-theory variables -- there are the following kinds of subsectors in field space: 

\begin{itemize}

    \item[-] The leading direction of the limit $\bm{v}_0 = \bm{e}_0$. 
    
    By construction, this selects the leading term of $\cV_X$, which depends on this field direction $M_0$ as $\cV_X \sim M_0^{w}$. As a consequence, $\cI_{00} \propto \cK_0^2$ and $\gamma_0 \sim \cO(1)$ \cite{Castellano:2024gwi}, so it belongs to the gravitational sector. More precisely, at the trajectory endpoint we have that $J \propto e_0^a \om_a$, so we can identify this $U(1)$ with the graviphoton. 

\item[-] Additional field directions $\bm{v}_A$ such that $\g_{\bm{v}_A} \sim \cO(1)$ asymptotically. 

\item[-] Field directions $\bm{v}_m$ with $\gamma_{\bm{v}_m} \to \infty$ along the limit. We dub this sector extended RFT.

\item[-] Field directions $\bm{v}_\mu$ such that $\gamma_{\bm{v}_\mu} \to \infty$ and $\cI_{\mu\mu} \to 0$ along the limit. This is the subsector of the extended RFT that goes to infinite coupling, and that we dub core RFT.
    
\end{itemize}

The precise definition for each of these subsectors depends on the scaling weight $w$ of the limit, and it is worked out in Appendix \ref{ap:asymptotic}. An important point is that, in the way that growth sectors have been defined, $M_0 \propto e^a_0$ is the only field direction that does not vanish along the trajectory. Another crucial point is that these $\bm{v}$'s, which are defined in terms of the triple intersection numbers $\cK_{abc}$ and the leading vector $\bm{e}_0$, form an integer basis of the string charge lattice. As such, they correspond to integrally-quantised $U(1)$'s that can be decoupled from each other if the appropriate kinematical conditions are met. 

By kinematical conditions we mean the standard field theory criterion that allows one to decouple a gauge sector with parametrically weaker gauge couplings from the rest. For instance, within the extended RFT sector, all $U(1)$'s that do not belong to the core RFT decouple from it along the trajectory. This is because when absorbing the gauge coupling in the field strengths, or equivalently when canonically normalising the scalar fields, the cubic Chern-Simons couplings get dressed with the different gauge couplings $g_a \simeq \cI_{aa}^{-1/2}$ as $\cK_{abc} \mapsto g_ag_bg_c \cK_{abc}$. As a result, all cubic couplings involving fields outside of the core RFT become negligible, and in this sense the rest of the extended RFT is seen as non-interacting and decoupled from the core RFT.  From the viewpoint of the core RFT, the rest of the extended RFT fields are then seen as non-dynamical parameters, that can only enter the core RFT through their vevs.

The separation between these two sectors also works because there are no parametrically large vevs for the scalars fields in this sector, such that they could generate a non-negligible kinetic mixing $\cI_{m\mu}  \simeq - \cK_{m\mu a}M^a$ between the two RFT sectors that one wants to decouple. Indeed, by the definition of core RFT given below \eqref{mK}, it turns out that such mixing terms do not depend on the field $M_0$ which, as pointed out above, is the only field direction that does not go to zero along the trajectories that we consider. As a consequence, the kinetic mixing terms between the core RFT and the rest of the extended RFT are suppressed with respect to the kinetic terms of the decoupling RFT sector, whose kinetic terms do depend on $M_0$. 

Let us now consider whether the gravitational sector of the 5d EFT also decouples from the RFT sector. We first look at the direction of the limit $\bm{v}_0$ which, in general, does not flow to weak coupling. Indeed, from the scaling $\cI_{00} \sim \phi^{\frac{2w}{3}-2}$ valid for EFT string limits, one can see that it even goes to strong coupling for infinite distance limits, where moreover we have that $M_0 \to \infty$. Hence, the above reasoning for the RFT sector cannot be applied in this case. Nevertheless, the decoupling with respect to the core RFT sector still occurs, because the definition below \eqref{mK} prevents any cubic Chern--Simons coupling mixing $\bm{v}_0$ with any core RFT direction. On the contrary, for the whole extended RFT there is no real decoupling at the level of interactions. Still, by the definition of RFT sector there is a suppressed kinetic mixing between $\bm{v}_0$ and the whole extended RFT, as one can check from the explicit expressions of Appendix \ref{ap:asymptotic}. 

Let us finally consider the decoupling of the field directions $\bm{v}_A$. In this case, it suffices to show that they do not flow to strong coupling along the limit to argue for their  decoupling with respect to the core RFT. One could again refer to the computations of Appendix \ref{ap:asymptotic} to check that this is the case for each kind of limit, but let us instead provide a more quantitative reasoning that is growth-sector independent. For this, we will use that when a $U(1)$ with $\gamma \sim \cO(1)$ flows to strong coupling, its magnetically charged BPS string charge becomes tensionless. In this way we can relate the decoupling statement to the spectrum of tensionless supergravity strings along the limit. We will run the analysis separately for each value of the scaling weight $w$.

For finite distance limits ($w=3$)  we sit at a finite-distance point in the K\"ahler cone, where all Nef divisors have finite volume and so all supergravity strings have finite tension. Their tension comes precisely from their coupling to the graviphoton, which is represented by $J \propto e_0^a \om_a$ or the Nef divisor $\cD_{\bm{e}_0}$. One can complete a divisor basis with a set of non-Nef divisors $\cD_{\bm{e}_m}$ such that ${\bf k}_a e^a_m = 0$,  corresponding to the extended RFT sector. It follows that there are are no vectors $\bm{v}_A$ for this class of limits, as one can check from \eqref{apgammaw3}, and so the decoupling follows trivially.

For decompactification limits ($w=2$) there is at least one additional gravitational sector besides $\bm{v}_0$, represented by the Kaluza--Klein $U(1)$, which always tends to zero coupling. Because the limit uplifts to a finite trajectory in 6d, by a similar reasoning as before one does not expect further gravitational sectors. Indeed, one can check that this KK sector corresponds to the direction $\bm{v}_E$ in \eqref{apvsw2}, and that the remaining sectors $\{\bm{v}_I, \bm{v}_\mu\}$ correspond to the extended RFT. In fact, since the latter come from a circle reduction of 6d fields at finite or infinite coupling, the direction $\bm{v}_E$ has a weaker coupling than all of them, and so one can again decouple this piece of the gravitational sector from the extended RFT. 

For emergent string limits ($w=1$) there should be a single supergravity string that becomes tensionless, or else there would be a contradiction with the Emergent String Conjecture \cite{Lee:2019wij}. Such a tensionless string is charged under the limit direction $\bm{v}_0$. All the remaining gravitational sectors, that correspond to $\bm{v}_I$ in \eqref{apvsw1}, must be weakly coupled, and contain those Kaluza--Klein $U(1)$'s that descend together with the emergent string \cite{Lee:2019wij,Kaufmann:2024gqo}. Finally, in this case the extended RFT and the core RFT sectors coincide, so one obtains a full decoupling between gravitational and RFT sectors.

To sum up, setting the graviphoton aside, we find a hierarchy of gauge couplings at any limit, that implies a nested decoupling of vector multiplet sectors at the level of interactions. The gravitational vector multiplets $\bm{v}_A$, which only exists for infinite distance limits, are at parametrically smaller coupling than the extended RFT sector, from which they decouple. Moreover, within the extended RFT the core RFT subsector is by definition at parametrically stronger coupling, which implements a further decoupling. When we include the graviphoton in the picture, we find that it only decouples from the core RFT, because the cubic interaction terms vanish identically. At the level of the extended RFT there is only a separation in the sense of kinetic mixing. More precisely, we find a block-diagonal structure at the level of the gauge kinetic matrix, with one block being the gravitational sector $\{\bm{v}_0, \bm{v}_A\}$, another one the core RFT $\{\bm{v}_\mu\}$, and the third one the rest of the extended RFT. The fact that one can neglect the kinetic mixing between decoupling sectors implies that one can replace the inverse $\cI^{\mu\nu} \simeq - \cK^{\mu\nu}$ restricted to the core RFT sector with the inverse of the restriction $\cI_{\mu\nu} \simeq - \cK_{\mu\nu}$ to this sector, and similarly for the extended RFT restriction $\cI_{mn} \simeq -\cK_{mn}$.\footnote{This feature, together with a charge-to-tension ratio with the largest divergence, were the criteria used in \cite{Castellano:2024gwi} to define a decoupling RFT. As can be verified with the computations of Appendix \ref{ap:asymptotic},   the core RFT always has the largest $\gamma_{\bm{p}}$. Hence, we see that in this setting one can describe a decoupling RFT with two different criteria.\label{ft:criteria}}

Finally, note that the above statements refer to decouplings at the level of interactions. In general, the subsectors that we have described can also couple to each other at the level of vevs of scalar fields. We say that the core RFT is fully decoupled from the other sectors when its couplings do not depend on their vevs. In the following, we will see that both the core RFT and its couplings at the level of vevs play a key role to describe scalar curvature divergences. 

\subsection{The rigid curvature}
\label{ss:rigid}

We now turn to analyse the r\^ole that the core and extended RFTs play in rigid limits with a diverging scalar curvature. For this, we first summarise some of the key results in Appendix \ref{ap:asymptotic}, which performs a detailed analysis of curvature divergences along limits of the form \eqref{growth}:

\begin{itemize}

    \item[-]  One can define the core RFT as follows. Consider the matrix
    \be
    {\bf K}_{ab} \equiv \cK_{abc}e_0^c\, , 
    \label{mK}
    \ee
    and compute its kernel. The core RFT is then described as the quotient space  $\ker' {\bf K} \equiv \ker {\bf K}/\langle \bm{e}_0 \rangle$ for $w=1$ limits,  and by $\ker' {\bf K} = \ker {\bf K}$ for $w=2,3$ limits. This is because $\ker {\bf K}$ contains the limit direction $\bm{e}_0$ in emergent string limits, which  corresponds to the graviphoton.

    \item[-] If $\ker' {\bf K} = 0$, no curvature divergence arises along the limit \eqref{growth}.  If $\ker' {\bf K} \neq 0$, divergences may appear with different powers of $\phi$. For EFT string limits with scaling weight $w$, the largest divergence is at most $\phi^w$.

    \item[-] To compute the divergence at this order, it suffices to consider the term
    \be
    R_{\rm div}^{\rm lead} = - \hat{\cK}^{\m\n} \hat{\cK}^{\r\s} \hat{\cK}^{\t\l} \cK_{\m\r[\t} \cK_{\n]\s\l}\, ,
    \label{Rlead}
    \ee
    where all indices belong to $\ker' {\bf K}$, and $\hat{\cK}^{\m\n}$ is the inverse of the matrix $\cK_{ab}$  restricted to the kernel indices. That is, it satisfies $\hat{\cK}^{\m\n} \cK_{\n \l} = \delta^\m_\l$, with $\nu \in \ker' {\bf K}$. The replacement $\cK^{\mu\nu} \to \hat{\cK}^{\mu\nu}$ signals the absence of kinetic mixing of the core RFT sector with the rest.

    \item[-] To compute milder curvature divergences, one must consider the extended RFT sector, which covers all the string ratios that diverge. The diverging piece of the curvature reads
    \be
    R_{\rm div} = - \hat{\cK}^{mn} \hat{\cK}^{rs} \hat{\cK}^{tl} \cK_{mr[t} \cK_{n]sl}\, ,
    \label{Rsublead}
    \ee
    where the indices $m$ now run over the extended RFT sector. Again $\hat{\cK}^{mn}$ is the inverse of the restriction of  $\cK_{ab}$ to the extended RFT sector: $\hat{\cK}^{mn} \cK_{n l} = \delta^m_l$. Its presence in \eqref{Rsublead} is due to the absence of kinetic mixing between the extended RFT and  gravitational sectors.

\end{itemize}

In summary, we have that $\ker' {\bf K}$ corresponds to a strong coupling rigid sector that can source a curvature divergence. Such a rigid theory is described by a prepotential of the form 
\be
\cF_{\rm rigid} = -\frac{1}{6} \cK_{abc} M^a M^b M^c \, .
\label{Frigid}
\ee
where the indices run over all vector multiplets. However, from the viewpoint of the core RFT, the coordinates $M^{a\neq\mu}$  that do not belong to $\ker' {\bf K}$ must be treated as non-dynamical parameters, which may nevertheless contribute to the rigid metric as constant shifts, since
\be
\cK_{\mu\nu} = \cK_{\mu\nu a} M^a\, .
\label{Kmunu}
\ee
The curvature that corresponds to this rigid metric is given by \eqref{Rlead}. The curvature \eqref{Rsublead}, which we also dub rigid curvature, is instead constructed from considering the rigid metric $\cK_{mn}$ and treating all fields of the extended RFT as dynamical, while  fields beyond the extended RFT are treated as non-dynamical parameters, contributing to the metric at most as constant shifts.

Notice that the leading piece of the rigid curvature goes like $g^6_{\rm rigid}$, where $g_{\rm rigid}$ is an average of gauge couplings of the core RFT. This quantity is of the same order as $\gamma_{\bm{q}}^2$, where $\bm{q}$ is an electric charge within the core RFT. As pointed out in footnote \ref{ft:criteria}, the core RFT  contains the largest ratios of the whole vector multiplet sector, which in turn suggests the following inequality
\be
R_{\rm div} \lesssim \gamma_{\rm max}^2\, ,
\label{Rgmax}
\ee
where $\g_{\rm max}$ is the largest charge-to-mass particle ratio along the limit. By the electro-magnetic duality relation  \eqref{pqdual} one expects a string to realise this maximal ratio as well, with the natural choice being a string wrapping a divisor within $\ker' {\bf K}$. In the examples of section \ref{s:examples} we will check this expectation as well as the inequality \eqref{Rgmax}, which is reminiscent of the upper bound found in \cite{Castellano:2024gwi} in the context of type II Calabi--Yau compactifications.  A related result found in \cite{Castellano:2024gwi} is that, whenever a curvature divergence occurs, $\ker' {\bf K}$ is generated by a set of effective divisors. Given the similarities between type IIA large-volume limits with curvature divergences and their M-theory counterparts, one expects the same result to apply in the present setup. In particular, we expect $\ker' {\bf K}$ to be generated by a set of effective, non-Nef divisors that collapse along the limit, giving rise to a set of non-critical, tensionless strings magnetically charged under the core RFT. Such tensionless strings are typically related to non-trivial rigid theories like SCFTs or Little String Theories (LSTs) that decouple from gravity. Moreover, from the results of \cite{Marchesano:2024tod}, one also expects such theories to appear as the UV rigid completion (UVRT) of the core RFT.

To gain further insight into this matter, one should provide some  geometric intuition behind the leading divergences contained in the core rigid curvature \eqref{Rlead} or, when absent, of the milder divergences within \eqref{Rsublead}. To simplify this task, we will consider setups in which the core RFT has a low dimension, more precisely when $r \equiv \dim (\ker' {\bf K}) = 1$ or $2$. This  restriction still allows to cover a large number of examples, like the ones considered in section \ref{s:examples}. 

\subsubsection*{One-dimensional core RFT}

In the case where $\ker' {\bf K}$ has a single dimension $\langle \bm{v}_{\mu}\rangle$, the core RFT curvature vanishes, and so does the leading term \eqref{Rlead} due to the index antisymmetrisation. One is then prompted to consider the extended RFT curvature, which can be rewritten as
\be
R_{\rm div} = \frac{1}{2} \cK_{mrt} \cK_{nsl} \hat{\cK}^{rs} \left( \hat{\cK}^{mt} \hat{\cK}^{nl} -\hat{\cK}^{mn}  \hat{\cK}^{tl}  \right)\, ,
 \label{alt}
\ee
with $m = (\mu, I)$. From this reordering, it is easy to see that $t \neq n$ and $m \neq l$ for the term to be non-vanishing. Additionally, the entries $\hat{\cK}^{\mu I}$ and $\hat{\cK}^{IJ}$ are always more suppressed than the core entries $\hat{\cK}^{\mu \nu}$, as follows from the results of Appendix \ref{ap:asymptotic}. The leading terms  then are 
\begin{align}
 \cK_{\mu \mu I} \cK_{\mu \mu J} \hat{\cK}^{\mu\mu} \left( \hat{\cK}^{\mu I} \hat{\cK}^{\mu J} - \hat{\cK}^{\mu\mu}  \hat{\cK}^{IJ}  \right) \simeq -\cK_{\mu \mu I} \cK_{\mu \mu J} \hat{\cK}^{\mu\mu} \hat{\cK}^{\mu\mu}  \hat{\cK}^{IJ} \, , \\
\cK_{\mu \mu \mu} \cK_{I \mu J} \hat{\cK}^{\mu\mu} \left( \hat{\cK}^{\mu \mu} \hat{\cK}^{IJ} - \hat{\cK}^{\mu I}  \hat{\cK}^{\mu J}  \right) \simeq \cK_{\mu \mu \mu} \cK_{I \mu J} \hat{\cK}^{\mu\mu} \hat{\cK}^{\mu\mu}  \hat{\cK}^{IJ} \, .
\end{align}
We finally obtain that the leading term in \eqref{Rsublead} reads
\be
R_{\rm div} = \frac{\hat{\cK}^{IJ}}{\cK_{\mu\mu}^{2}} \left( \cK_{\mu \mu \mu} \cK_{\mu IJ} - \cK_{\mu \mu I} \cK_{\mu \mu J} \right) + \dots
\label{Rdivr1}
\ee

\subsubsection*{Two-dimensional core RFT}

With two dimensions, the core RFT curvature \eqref{Rlead} can be non-vanishing. Considering \eqref{alt} with the sum over indices restricted to core RFT ones $\mu, \nu =1,2$, one finds that the factor in parenthesis is given by $(\det \cK_{\mu\nu})^{-1}$, and factors out from the sum. Summing up the additional terms one finds 
\be
R_{\rm div}^{\rm lead} = 
\frac{\det {\bf M}}{\left(\det \CK_{\mu\nu}\right)^2} \, ,
\label{Rdivr2}
\ee
where
\be
{\bf M} = 
\begin{pmatrix}
    \cK_{111} & \cK_{112} & \cK_{11}   \\
     \cK_{121} & \cK_{122} & \cK_{12} \\
     \cK_{221} & \cK_{222} & \cK_{22} 
\end{pmatrix} \, .
\label{matrixM}
\ee
It follows from \eqref{Kmunu} that, to have $\det {\bf M} \neq 0$, one must have $\cK_{\mu\nu a} \neq 0$, for some $a \notin \ker' {\bf K}$.

We thus find that, in both cases, the core RFT sector must couple to some other sector through the Chern--Simons terms, for the leading divergence to be non-vanishing. To give a more precise statement, and to furnish both a geometrical and physical interpretation of what these divergences mean, it is necessary to analyse them for each class of limits separately. 


\section{Finite distance limits}
\label{s:w3}

Let us consider those divergences of the scalar curvature that can occur along $w=3$ limits, which are finite-distance trajectories of the form \eqref{Mtrajw3}. It follows from the analysis in \cite{Marchesano:2023thx} that when $\ker {\bf K}$ is non-trivial, at $M^a = e_0^a ({\bf k}/6)^{1/3}$ there is one or several contractible divisors that collapse to a point. Such  divisors are non-Nef, effective and generalised del Pezzo surfaces, with a non-trivial triple intersection number, and we have one of them per independent element of $\ker {\bf K}$. Physically, we are at a strong coupling singularity in moduli space where a rank $r \equiv \dim (\ker {\bf K})$ 5d SCFT  that appears  \cite{Witten:1996qb,Intriligator:1997pq,Jefferson:2018irk}, and  several infinite towers of particles become massless. Moreover, at this point $r$ non-critical strings become tensionless, and the gauge couplings of the $r$  $U(1)$'s under which they are magnetically charged become infinite in 5d Planck units. In particular, for EFT string limits the gauge kinetic matrix elements look like\footnote{To simplify the discussion, we have chosen a basis of divisors such that $\cK_{00I} = 0$ as in \cite[eq.(2.12)]{Alim:2021vhs}].}
\be
\begin{split}
\cI_{00} = 3  \left(\frac{\bf k}{6}\right)^{2/3}  \equiv c \, , & \qquad  \cI_{IJ} = - \left(\frac{6}{\bf k}\right)^{1/3} \cK_{0IJ} \equiv b_{IJ}\, , 
\\
\cI_{\mu\nu} \simeq -\cK_{\mu\nu} \sim \phi^{-1}\, , & \qquad  \cI_{\mu I} \simeq -\cK_{\mu I}\sim \phi^{-1}\, ,
\end{split}
\label{Iw3}
\ee
where the index $a =0$ represents the direction of the limit $\bm{e}_0$, the indices $\mu, \nu$ the directions of the core RFT and $I, J$ the directions of the extended RFT outside of the core. One can see that these last set of elements correspond to gauge interactions with a finite gauge coupling, but that nevertheless we have $\gamma_I  \to \infty$ because $\cK_I \to 0$. It is however important to stress that one cannot interpret $\cK_I/2$ as the tension of a 5d string, since the divisors ${\cal D}_I$ that correspond to this sector are generically not effective. Finally, in the case of general growth sectors, one has to replace $\phi^{-1} \to \phi^{-\eta}$ in the above scalings, where $\eta  > 0$ depends on the growth factors $\gamma_i$ that appear in \eqref{growth}. Here we assume a universal growth factor $\gamma = \gamma_i$, $\forall i$, so that $\eta = 1 - \gamma$.

With these observations, one can turn to evaluate the explicit expressions for the leading terms of the curvature divergence, obtained for the cases $r=1,2$. In the case of a one-dimensional core RFT, we have that \eqref{Rdivr1} becomes
\be
R_{\rm div} = \phi^{2\eta}  \kappa_\mu^{-2}\, b^{IJ}\left(\cK_{\mu \mu I} \cK_{\mu \mu J} - \cK_{\mu \mu \mu} \cK_{\mu IJ}\right) + \dots
\label{Rdivr1w3}
\ee
where $\kappa_\mu \simeq (6/{\bf k})^{1/3} (\cK_{\mu\mu\mu} + \sum_I\cK_{\mu\mu I})$ is in general non-vanishing. The coefficient that arises from the sum over $I,J$ is non-negative. It vanishes if there is no intersection between the divisors ${\cal D}_I$ and the core RFT divisor ${\cal D}_\mu$, and it is positive otherwise. In fact, if $\cK_{\mu IJ} = \cK_{\mu \mu I} = 0$, $\forall I,J$, it is easy to see that \eqref{Rsublead} can only provide a finite term at the endpoint of the trajectory, and thus there is no divergence. 

To see that in any other case the curvature diverges, recall that in this setup $\cK_{\mu\mu\mu} \neq 0$. Because it is a contractible surface, and growing its volume reduces the CY volume, in the convention in which $M^\mu >0$ we have that  $\cK_{\mu\mu\mu} < 0$ and ${\cal D}_\mu$ is anti-effective. Applying the Hodge index Theorem (see e.g. \cite[Theorem A.32]{Katz:2020ewz}) to this convention we see that the matrix
\be
({\bf K}_{\mu})_{ab} = \cK_{\mu ab}
\label{Kmu}
\ee
has only one negative eigenvalue, some zero eigenvalues (like the trajectory divisor ${\cal D}_{\bm{e}_0}$ and additional divisors ${\cal D}_I$ that do not intersect ${\cal D}_\mu$), and the rest are positive. Divisors of the form
\be
{\cal D}'_I = -\cK_{\mu\mu\mu} {\cal D}_I + \cK_{\mu\mu I} {\cal D}_\mu\, ,
\label{DIprime}
\ee
are always orthogonal to ${\cal D}_\mu$ with respect to the Lorentzian $(-,+,+,+,\dots)$  metric \eqref{Kmu}. Since ${\cal D}_\mu$ is timelike, one then deduces that the ${\cal D}'_I$ must have positive or vanishing norm, and since
\be
{\cal D}_\mu \cdot {\cal D}'_I \cdot {\cal D}'_J = \cK_{\mu\mu\mu} \left(\cK_{\mu \mu \mu} \cK_{\mu IJ} - \cK_{\mu \mu I} \cK_{\mu \mu J} \right)\, ,
\label{Dprimenorm}
\ee
it follows that the tensor in \eqref{Rdivr1w3} is positive semidefinite. Finally, since $b_{IJ}$ is positive definite, their contraction gives a positive number whenever \eqref{Dprimenorm} does not vanish identically.

For the case $r=2$ a cubic divergence is generated unless $\det {\bf M}$ identically vanishes, because $\cK_{\mu\nu}(\det \cK_{\mu\nu})^{-2} \sim \phi^{3\eta}$ along the trajectory \eqref{Mtrajw3}. For $w=3$ setups we have that $\cK_{\mu\nu} =\cK_{\mu\nu \rho} M^\rho + \cK_{\mu\nu I} M^I$, but the sum over core RFT indices does not contribute to the determinant. Hence, to have a leading divergence we need that a divisor of the sector ${\cal D}_I$ has a non-trivial intersection with some of the divisors in $\ker {\bf K}$. More precisely, $\det {\bf M} \neq 0$ if and only if the intersection curves within the core RFT divisors
\be
{\cal D}_1 \cdot {\cal D}_1\, , \qquad {\cal D}_1 \cdot {\cal D}_2\, , \qquad {\cal D}_2 \cdot {\cal D}_2\, , 
\label{interD}
\ee
are linearly independent as vectors of the Mori cone. If this is not the case, and $\CK_{\mu\nu I} =0$ $\forall I$, divergences can be generated at a subleading order via the term 
\be
R_{\rm div} = - 2 \hat{\cK}^{\mu\nu} \cK_{\mu IJ} b^{IJ} \p_\nu \log (\det \cK_{\mu\nu}) + \dots
\ee
namely a product of two non-trivial vectors, that in general does not vanish. If, on the other hand, also $\cK_{\mu IJ} = 0$ $\forall I,J$ the core RFT decouples from the rest of the extended RFT, and no divergence is generated. 

Finally, following \cite{Marchesano:2024tod} let us compare the scalar curvature divergence with the behaviour of the quotient of scales $\Lambda_{\rm wgc}/\Lambda_{\rm sp}$ near the singularity, where  $\Lambda_{\rm wgc}^2 \equiv g_{\rm RFT}^2 M_{\rm P}^{3} \simeq \cI^{\mu\nu}M_{\rm P}^2$ is the core RFT cut-off predicted by the magnetic WGC, and $\Lambda_{\rm sp}^2 \simeq M_{\rm P}^2 \left(\int_X c_2(X) \wedge J\right)^{-1}$ is the species scale estimate of \cite{vandeHeisteeg:2023dlw}. It is easy to see that in the case at hand 
\be
\frac{\Lambda_{\rm wgc}}{\Lambda_{\rm sp}} \simeq \phi^{\frac{\eta}{2}}\, \implies \,R_{\rm div} \sim \left(\frac{\Lambda_{\rm wgc}}{\Lambda_{\rm sp}}\right)^{2\nu}\, ,
\label{Rwgcsp}
\ee
with $\nu = 3$ for leading divergences and $\nu =2$ for subleading ones. 

\subsubsection*{Physical interpretation}

As it follows from the general discussion in the previous section, a leading curvature divergence of the form $R_{\rm div} \sim \phi^{3\eta}$ (i.e. $\nu =3$ in \eqref{Rwgcsp}) corresponds to a core RFT with a non-vanishing curvature \eqref{Rlead}. If the core RFT has a flat metric, one may only have a subleading divergence ($\nu=2$ in \eqref{Rwgcsp}) or no divergence at all. However, from this perspective it is not obvious how to give a physical meaning between these last two cases. 

It turns out that one can link the divergences in the scalar curvature with how the core RFT couples (at the vev level) with the other vector multiplets, that is with the divisors ${\cal D}_I$ of the extended RFT. Recall that, at the trajectory endpoint, the core RFT becomes a 5d SCFT, and the additional fields $M^I$ in the extended RFT non-dynamical mass parameters \cite{Jefferson:2018irk}. Our previous analysis for rank 1 and 2 core RFT implies that, if the core RFT is fully decoupled from the remaining vector multiplets, then no curvature divergence is generated. By fully decoupled we mean that the the cubic Chern--Simons terms mixing the core RFT with any other sector vanish identically. As a result the 5d SCFT is independent of the mass parameters $M^I$ along its Coulomb branch. Neither the gauge kinetic functions $\cI_{\mu\nu}$ nor the tension $T_\mu/M_{\rm P}^2$ of the strings charged under the SCFT depend on the $M^I$. Geometrically, the absence of divergence constrains the curves within the contractible divisors $\cD_\mu$. A full decoupling amounts to intersections ${\cal D}_\mu \cdot {\cal D}_I$ that are homologically trivial, and so they do not support BPS particles. The intersections ${\cal D}_\mu \cdot {\cal D}_\nu$, which do correspond to non-trivial  classes, only intersect with the core RFT divisors. These curves are moreover effective and, due to the properties of del Pezzo surfaces, one expects to be able to perform multi-wrappings on them \cite{Witten:1996qb}. Physically, the towers of BPS particles that become massless at the 5d SCFT point are only charged under the core RFT gauge group.

 In the case where there is a divergence, at least some of the intersections ${\cal D}_\mu \cdot {\cal D}_I$ are non-trivial.  These typically correspond to effective curves and, in fact, one can again associate them to towers of particles that become massless at the 5d SCFT point, but now displaying a wider spectrum of charges involving also the extension of the RFT sector by the divisors ${\cal D}_I$. This coupling to the extended RFT is also manifest in terms of the tensions of the core RFT  strings, which now depend on the mass parameters. Finally, if the gauge kinetic matrix $\cI_{\mu\nu}$ always depends on the mass parameters we have a maximal curvature divergence with $\nu=3$ in \eqref{Rwgcsp}, while if some redefinition of the $M^I$ removes this dependence we have that $\nu=2$ instead. 

To illustrate this picture, let us consider a rank 1 core RFT decoupled from the extended RFT sector, that then undergoes a flop transition. Before the flop we have the rigid prepotential 
\be
\cF_{\rm rigid} = - \frac{1}{6} \cK_{\mu\mu\mu}( M^\mu)^3 + \oh b_{IJ} M^I M^J - \frac{1}{6} \cK_{IJK} M^IM^JM^K\, ,
\ee
where $b_{IJ}$ is defined as in \eqref{Iw3}. Only the first term hosts dynamical fields from the viewpoint of the core RFT, yielding a vanishing curvature for this sector. As $M^\mu \to 0$ the rigid string made up from an M5 brane wrapping the divisor $\cD_{\mu}$ becomes tensionless, reaching a 5d SCFT point. The same happens to the tower of particles that come from wrapping an M2-brane on the curve $\cD_\mu \cdot \cD_\mu$ or its multiwrappings, as expected from electric-magnetic duality. Indeed, notice that the relation \eqref{pqdual} reduces in the rigid limit to
\be
q_a \propto \CK_{ab}p^b\, .
\ee
In this case, it maps the RFT string charge $p^\mu$ to $q_\mu$, which corresponds to the class $\cD_\mu \cdot \cD_\mu$.

Let us now perform a flop transition along a curve ${\cal C}_f = f_\mu {\cal C}^\mu + f_I {\cal C}^I$, where we are considering a dual basis of curves such that ${\cal C}^a \cdot {\cal D}_b = \delta^a_b$. The change in the prepotential after the flop is 
\be
\Delta \cF_{\rm rigid} =  \frac{n_{{\cal C}_f}}{6} \left( f_\mu M^\mu + f_I M^I\right)^3 \, ,
\ee
where $n_{{\cal C}_f}$ is the genus zero GV invariant of the flop curve, and by construction we must have $f_I \neq0$. As a result, the tension of the core RFT string now reads
\be
T_\mu/ M_{\rm P}^2  \sim  - \frac{1}{4}  \cK_{\mu\mu\mu} (M^\mu)^2  +  \frac{1}{4} n_{{\cal C}_f} f_\mu \left( f_\mu M^\mu + f_I M^I\right)^2\ , 
\ee
where recall that $\cK_{\mu\mu\mu} < 0$. The metric and gauge kinetic term for the core RFT reads
\be
\cI_{\mu\mu} \simeq -\cK_{\mu\mu} = -\cK_{\mu\mu\mu} M^\mu +  n_{{\cal C}_f} f_\mu^2 \left( f_\mu M^\mu + f_I M^I\right)\, .
\ee
The flop modifies the core RFT CS couplings, together with its kinetic terms and string tension. When  $n_{{\cal C}_f} f_\mu < 0$ one can see that this also implies a topology change on the divisor $\cD_\mu$, in which the flop curve has been blown up. If the blown-up curve can be understood as the intersection of two effective divisors, this implies that $\cD_\mu$ hosts a richer set of non-trivial curves in terms of the CY homology. In physics terms, we find a tower of particles arranged in a two-dimensional cone of charges, all of them becoming massless as we approach the SCFT point. 

\begin{figure}[ht!]
\hspace{0cm}
\includegraphics[width=16cm]{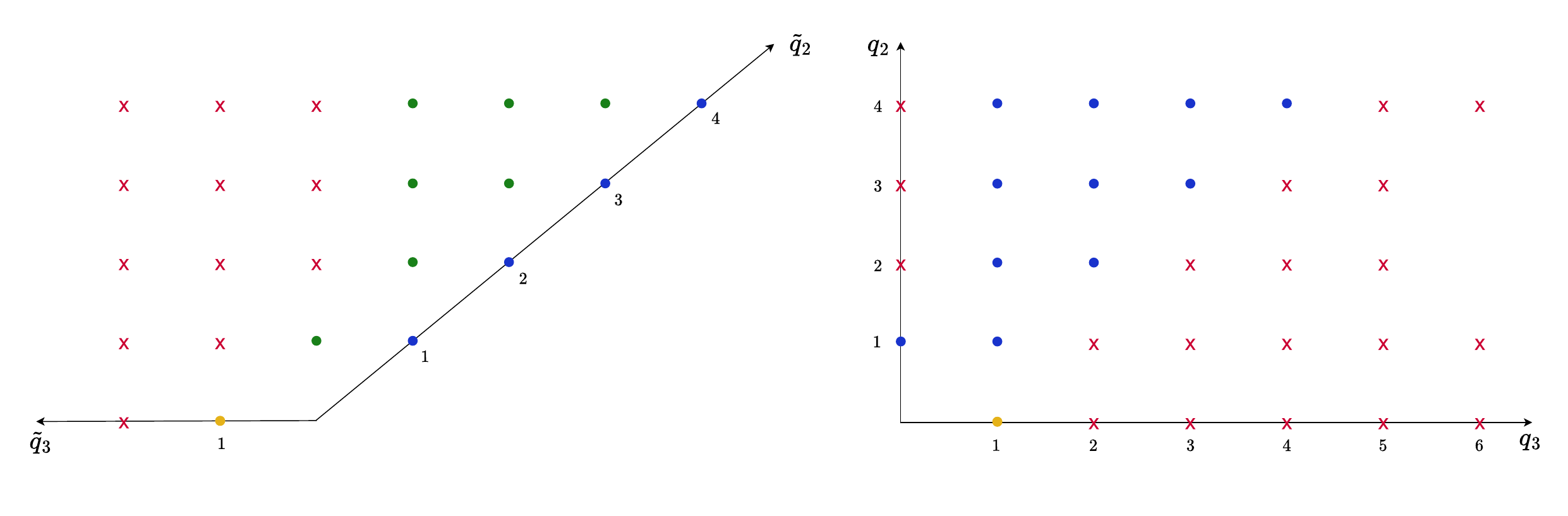}
\caption{Genus-zero GV invariants of the KMV conifold in the $\mathbb{P}^2$ (left) and the $\mathbb{F}_1$ phase (right), excluding the direction of the elliptic fibre. The dots represent non-zero GV invariants, while the crosses correspond to vanishing ones \cite{Alim:2021vhs}.  The yellow dots represent the flop curve in each phase. The blue dots represent the curves contained within $\mathbb{P}^2$ and $\mathbb{F}_1$, respectively, and which go to zero area  when the base collapses. 
\label{fig:GV_KMV}}
\end{figure}

This is indeed what happens in the case of the KMV conifold \cite{Klemm:1996hh}. The above discussion describes the flop transition in the direction that changes the topology of the divisor $\cD_\mu$ from $\mathbb{P}^2$ to $\mathbb{F}_1$,\footnote{Notice that in the literature the phase with $\cD_\mu = \mathbb{P}^2$ is dubbed the flopped phase and the one with $\cD_\mu = \mathbb{F}_1$ the unflopped one, somewhat opposite to the perspective taken in our discussion.} with parameters $n_{{\cal C}_f} =1$, $f_\mu =-1$ and $f_I = -8$. The distribution of effective curves with non-trivial GV invariants before and after the flop is illustrated in figure \ref{fig:GV_KMV}. When $\cD_\mu$ is given by a $\mathbb{P}^2$, all the curves within it are multiwrappings of its self-intersection, and M2-branes wrapped on them are only charged under the gauge group of the core RFT. On the other side of the flop transition, where $\cD_\mu$ has the topology of a $\mathbb{F}_1$, the M2-branes wrapping effective curves within $\mathbb{F}_1$ arrange into a two-dimensional cone that also has non-trivial charges under the $U(1)$ that corresponds to $\cD_I$. Physically, they can be seen as bound states of the tower $n \cD_\mu \cdot \cD_\mu$ and the hyper that mediates the flop transition. As a result, their masses depend on the mass of such a hyper which, from the viewpoint of the core RFT, is understood as a non-dynamical mass parameter \cite{Jefferson:2018irk}. Notice that the tension of the core RFT string also depends on such a parameter. 
The second effect observed above, namely that the gauge kinetic term of the core RFT is also modified, is however more subtle. In particular, one may choose the freedom that one has in the definition of the extended RFT sector to remove the dependence on the mass parameter $M^I$. Indeed, since in general the divisors $\cD_I$ are defined up to elements of $\ker' {\bf K}$, one may replace $\cD_I$ by the combination \eqref{DIprime} which by construction does not intersect $\cD_\mu \cdot \cD_\mu$. In terms of $\cD_I'$ and its K\"ahler parameter $M'^I$ the gauge kinetic term of the core RFT reads
\be
\cI_{\mu\mu} \simeq -\left( \cK_{\mu\mu\mu} - n_{{\cal C}_f} f_\m^3 \right) M^\mu  \, ,
\ee
while the tension of the core RFT string is
\be
T_\mu/ M_{\rm P}^2  \sim  - \frac{1}{4} \left( \cK_{\mu\mu\mu} - n_{{\cal C}_f} f_\m^3 \right) (M^\mu)^2  + \frac{1}{4} \left( n_{{\cal C}_f} f_\m f_I^2 \frac{\cK_{\m\m\m}}{\cK_{\m\m\m} - n_{{\cal C}_f} f_\m^3} \right) (M'^I)^2 \, . 
\ee
\label{tensionflop}
We thus see that the dependence of the kinetic terms on the mass parameters can be removed by an appropriate choice of basis, while the same is not true for the tension of the string and the mass of the particles charged under the core RFT. 

Remarkably, this difference is captured by the degree of divergence of the curvature. As mentioned above, when the rank of the core RFT equals one, its rigid curvature automatically vanishes, and one may only have subleading divergences via \eqref{Rdivr1w3}. In the case of a rank-two core RFT one can have a leading divergence if the determinant of the matrix \eqref{matrixM} does not vanish, where in the last column (minus) the elements of the kinetic matrix appear. When $\det {\bf M} =0$ this column is a linear combination of the other two, which means that by shifting the divisors $\cD_I$ with elements of $\ker {\bf K}$ one can remove all mass parameters from the kinetic terms. Contrarily, when $\det {\bf M} \neq 0$ one cannot remove such a dependence in any base, while simultaneously one obtains a maximal divergence of the scalar curvature. 

To sum up, we have analysed a set of finite-distance trajectories whose endpoint is a 5d SCFT of rank 1 or 2. The moduli space curvature may diverge or not at such a point, depending on how the 5d SCFT couples via vevs to the rest of the vector multiplet sector. When a divergence is present, some cubic CS terms mix the core RFT with other vector multiplets, and some towers of particles that become massless at the strong coupling singularity necessarily have non-trivial charges under them. This is in contrast with the string magnetic charges, that by construction only lie within the 5d SCFT. As a result the tension of core RFT strings and their dual particles depend on the K\"ahler moduli that are seen as mass parameters form the core RFT viewpoint. The dependence of the gauge kinetic matrix on such mass parameters, however, is only physical if and only if the scalar curvature displays a maximal divergence.  It would be very interesting to see if this picture holds for  5d SCFTs of higher rank, or if new features arise in those cases.


\section{Infinite distance limits}
\label{s:w21}

Having analysed when curvature divergences are present at finite-distance boundaries with a 5d SCFT of rank $r \leq 2$, we turn to study the analogous problem for boundaries at infinite distance in 5d vector moduli space. As pointed out in \cite{Lee:2019wij}, there are essentially two kinds of infinite-distance limits in this context, given by decompactification limits to 6d ($w=2$) and emergent string limits ($w=1$). We analyse each of this class of limits in turn, trying to extract the physical and geometric meaning behind a curvature divergence. In the case of decompactification limits, we obtain a picture that is somewhat similar to the one developed in \cite{Marchesano:2024tod} for decompactification limits of type IIA to F-theory. Indeed, the presence and physical interpretation of a curvature divergence depends heavily on the kind of shrinking divisors that correspond to the core RFT, and more precisely on their geometry from the viewpoint of the elliptic fibration of the Calabi--Yau. Certain divisors, like fibral divisors, give rise to 5d BPS strings whose tension is below the (squared) 6d KK scale of the limit, and in this sense the physics of curvature divergences is the same as for finite-distance boundaries. When the 5d string tensions are above this scale, as it happens for vertical and exceptional divisors, then the core RFT flows to a 6d SCFT in the UV, and the curvature divergence is instead linked to the presence of a non-Abelian gauge group.

\subsection{Decompactification limits}
\label{ss:w2}

Let us now consider the asymptotic behaviour of the moduli space scalar curvature in $w=2$ limits, which for EFT string limits are of the form
\be
M^a(\phi) =\frac{ t_0^a + e_0^a \phi}{ \left(\frac{1}{2} {\bf k}_a t_0^a\phi^2\right)^{1/3}} \left[ 1 + \cO(\phi^{-1}) \right] \, .
\label{Mtrajw2}
\ee
Physically, these correspond to infinite distance limits in which one dimension is decompactified. In the simplest cases, the resulting 6d EFT can be obtained from F-theory compactified on the elliptically fibered Calabi--Yau $X$, and the above 5d  trajectory can be mapped to a finite-distance trajectory in the moduli space of this 6d uplift. 

\subsubsection*{Smooth fibrations}

Let us first consider the case where $X$ is a smooth elliptic fibration, with base $B_2$. The triple intersection numbers in a Nef basis are  
\be
\cK_{EEE} = \eta_{\a\b}c_1^\a c_1^\b\, , \qquad \CK_{EE\a} = \eta_{\a\b} c_1^\b\, , \qquad \CK_{E\a\b} = \eta_{\a\b}\, , \qquad \CK_{\a\b\g} = 0\, ,
\ee
where ${\cal D}_E$ is dual to the elliptic fibre, and ${\cal D}_\a = \pi^*(\om_\a^{\rm b})$ is a basis of vertical divisors. In addition $c_1(B_2) = c_1^\a \om_\a^{\rm b}$ and $\eta_{\a\b}$ is the intersection matrix of the curves ${\cal C}_\a = B_2 \cdot {\cal D}_\a$ on $B_2$. A $w=2$ limit is obtained by choosing a vector $e_0^a=(0,e^\a)$, such that $\eta_{\bm{e}} \equiv \eta_{\a\b}e^\a e^\b > 0$ and then the trajectory \eqref{Mtrajw2}. This limit has a $\ker {\bf K}$ given by those divisors ${\cal D}_\mu = f_\mu^\a {\cal D}_\a$ such that $\eta_{\a\b} e^\a f^\b_\mu = 0$. Following section 4.3 of \cite{Marchesano:2024tod}, one sees that this trajectory uplifts to the following one in the tensor branch of the F-theory uplift
\be
j^\a = \frac{1}{\sqrt{\eta_{\bm{e}}}} \left[ e^\a + \phi^{-1} \left( t^\a_0 + \frac{1}{2} c_1^\a t^E_0 - \frac{1}{2\eta_{\bm{e}}} e^\a \eta_{\g\del} e^\g (2t^\del_0 + c_1^\del t^E_0)\right) \right] + \cO \left(\phi^{-2}\right) \, ,
\label{Ftrajw2}
\ee
where the F-theory variables are related to the M-theory ones by
\be
j^\a = \sqrt{\frac{M^E}{2}} \left( M^\a + \frac{1}{2} c_1^\a M^E \right)\, .
\ee
In the F-theory frame, the core RFT sector corresponds to the set of contractible curves ${\cal C}_\mu = f^\a_\mu {\cal C}_\a$ and the associated field directions $j^\mu$ in the tensor branch. A D3-brane wrapping a curve ${\cal C}_{\mu}$ will result in a 6d string whose tension goes like $T_\mu \simeq \eta_{\a\b} j^\a f_\mu^\b \sim \phi^{-1}$ in 6d Planck units, signaling that we are approaching a 6d SCFT regime. However, the  tensor branch metric
\be
g_{\a\b} = 2 j_\a j_\b - \eta_{\a\b}\, ,
\ee
where $j_\a = \eta_{\a\b}j^\b$, reduces to $c_{\mu\nu} = - \eta_{\a\b} f_\mu^\a f_\nu^\b$ for the core RFT sector. That is, it asymptotes to a flat, non-singular metric, dubbed (minus) the pairing matrix in the 6d SCFT literature \cite{Heckman:2018jxk}.

For this subclass of $w=2$ limits, one can easily see that the 5d scalar curvature cannot diverge along the trajectory \eqref{Mtrajw2}. Indeed, due to the condition $\CK_{\a\b\g} = 0$ the core RFT curvature \eqref{Rlead} vanishes identically. Moreover, the divisors ${\cal D}_I$ that define the extension of RFT are only present in non-smooth fibrations, which implies that \eqref{Rsublead} reduces to \eqref{Rlead} and no divergence can be generated whatsoever. One can compare this result with the one obtained in \cite{Marchesano:2023thx,Marchesano:2024tod} where it was found that, for  $w=2$ 4d limits \eqref{growth} based on smooth elliptic fibrations, curvature divergences can only be sourced by world-sheet instantons. The fact that in the  5d setup no divergence can generated is in agreement with our general expectations that such instanton effects are absent in the present M-theory frame (cf. footnote \ref{ft:instantons}). 

\subsubsection*{Vertical divisors}

In fact, the analysis in \cite{Marchesano:2023thx,Marchesano:2024tod} points to a more general result. The 5d curvature should also be absent of divergences along $w=2$ limits built from a Calabi--Yau $X$ that is a non-smooth elliptic fibration, provided that $\ker {\bf K}$ only contains vertical divisors. To arrive to this result in the present context, let us consider a flat elliptic fibration with at least one section, which we identify with the zero section $\cD_E$. Then by  \cite[eq.(2.17)]{Grimm:2013oga}, we have that the  triple intersection numbers follow the following structure
\be     \label{inters_vertical}
\cD_\a \cdot \cD_\b \cdot \cD_\g = \cD_\a \cdot \cD_\b \cdot \cD_I = \cD_\a \cdot \cD_\b \cdot \cD_m = 0 \, , \qquad \cD_\a \cdot \cD_\b \cdot \cD_E = \eta_{\a\b} \, ,
\ee
where $\cD_\a$ denote vertical divisors,  $\cD_m$ the images under the Shioda map of possible extra sections $\s_m$ and  $\cD_I$ represent exceptional divisors. In terms of the notation of Appendix \ref{ap:asymptotic}, $\cD_E$ corresponds to the vector $\bm{v}_E$, one of the vertical divisors $\cD_{e_0}$ corresponds to the leading limit direction $\bm{v}_e$, and the rest to extended RFT subspace $\langle \bm{v}_I \rangle \bigoplus \ker {\bf K}$. By assumption, $\ker {\bf K}$ only contains vertical divisors, which means that the leading divergence \eqref{Rlead} vanishes identically. This only leaves the next-to-leading order divergences contained in \eqref{Rsublead}, which in the case of EFT string limits is linear in $\phi$. However, the structure \eqref{inters_vertical} implies that also such subleading divergences vanish. Indeed, in the case of a one-dimensional core RFT, such divergences are given by the leading term in \eqref{Rdivr1}, which vanishes because $\cK_{\mu\mu\mu} = \cK_{\mu\mu I} = 0$. More generally, one needs to consider terms of the form 
\be
\cK^{IJ} \cK^{\m\n} \cK^{\r\s} \left( 2 \cK_{I\m [\r} \cK_{J]\n\s} + \cK_{I\m [\r} \cK_{\n]J\s} \right) \, ,
\ee
which also vanish due to \eqref{inters_vertical}. Hence in this case the extended RFT curvature \eqref{Rsublead} can at most give a constant contribution.

\subsubsection*{Connection to F-theory}

\begin{figure}[ht!]
\begin{center}
\includegraphics[width=14cm]{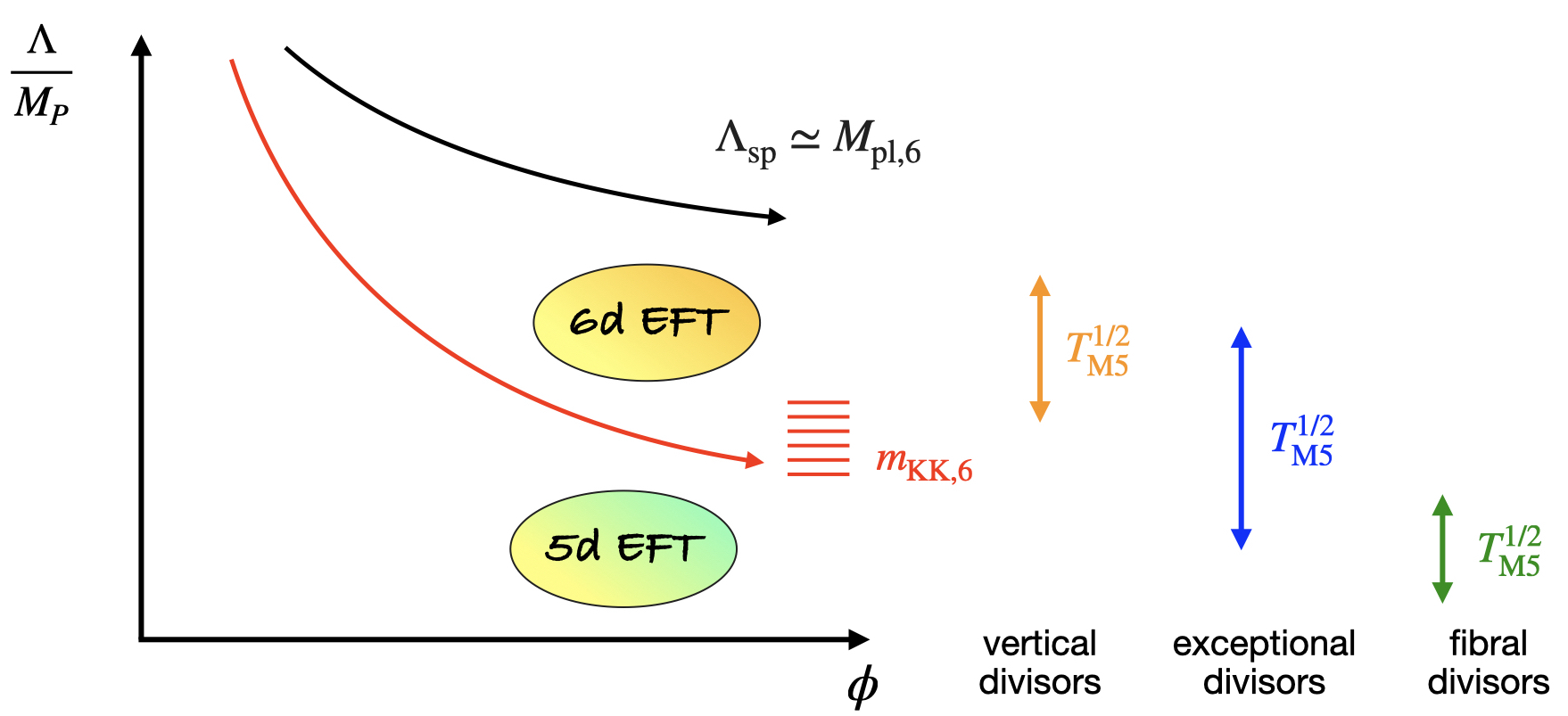}
\caption{String tensions coming from M5-brane wrapping different kinds of divisors within the kernel of $w=2$ limits, compared to the 6d Kaluza--Klein scale.  
\label{fig:tensions}}
\end{center}
\end{figure}

As it turns out, one could have anticipated both of the above results on physical grounds, by adapting the picture obtained in \cite{Marchesano:2024tod} for $w=2$ limits to the present setting. The key observation is that $w=2$ rigid limits can be classified in terms of the kind of divisors ${\cal D}_\mu$ that belong to $\ker {\bf K}$. This is because, when wrapping an M5-brane on an effective divisor within $\ker {\bf K}$, the asymptotic relation between the string tension and the 6d (squared) KK scale  depends on the geometry of the divisor and the growth sector that one considers, as illustrated in figure \ref{fig:tensions}. 
M5-branes wrapping vertical divisors of $\ker {\bf K}$ give rise to strings whose tension always lie at or above the $m_{\rm KK,6}^2$. In contrast, fibral divisors give rise to strings that always lie below $m_{\rm KK,6}^2$, while exceptional divisors provide strings that can lie either above or below. Finally, as argued in \cite{Marchesano:2024tod}, rational sections cannot belong to the core RFT.

The importance of this observation lies in that, when all string tensions lie above $m_{\rm KK,6}^2$, one can integrate these strings out and conclude that the EFT description at $m_{\rm KK,6}$ is nothing but the 6d supergravity obtained upon compactifying F-theory on $X$. Then, to obtain the 5d EFT whose moduli space curvature we want to compute, one only needs to perform a circle reduction of this 6d EFT. This lesson also applies to the subsector of the 6d EFT that descends to the 5d RFT. As mentioned above, in the case of smooth elliptic fibrations the 5d core RFT uplifts to a subsector of the tensor branch with a flat metric $c_{\mu\nu}$, that results in a 6d SCFT at the endpoint of the trajectory \eqref{Ftrajw2}. Performing a circle reduction of this metric gives a 5d flat metric as well, and that is why the core RFT curvature vanishes. For non-smooth elliptic fibrations where $\ker {\bf K}$ only contains vertical divisors, one can apply the same principle. The difference is that now the vectors $\bm{v}_I$ of the 5d extended RFT uplift to $n_I$ Abelian and non-Abelian 6d gauge vectors, with gauge couplings of the form
\be
j^\a \cK_{\a IJ}   F^I \wedge * F^J\, .
\label{Fgaugecoup}
\ee
The fact that the corresponding divisors do not belong to $\ker {\bf K}$ implies that their gauge couplings remain finite at the endpoint of \eqref{Ftrajw2}. Upon reduction on a circle these gauge bosons become $n_I$ axion fields, so the metric of the 5d extended RFT corresponds to a $T^{n_I}$ fibration over a flat space. The fact that the gauge couplings in 6d are finite implies that the fibration does not degenerate at the endpoint of the $w=2$ limit, which in turns means that we have a regular rigid metric at this point, and so the curvature \eqref{Rsublead} does not diverge.

When some of the string tensions lie below $m_{\rm KK,6}^2$,  the above strategy cannot be bluntly applied. Indeed, the 6d EFT that one recovers at $m_{\rm KK,6}$ is likely not to correspond to the IR limit of F-theory compactified on $X$ because there is a BPS string (and possibly some particles) that should not be integrated out at this energy scale. In this case, to have an accurate description of the limit one should consider F-theory compactified on $S^1 \times X$, or equivalently M-theory compactified on $X$. As in \cite{Marchesano:2024tod}, when a BPS string tension or particle mass lies at or below the 6d KK scale $m_{\rm KK,6} \simeq R_{S^1}^{-1}$, one should integrate it out as a 5d object. This results in a different 5d EFT in the IR compared to what one would have obtained if the object was integrated out in 6d, and gives a rationale of why the physics of infinite distance limits depend on which growth sector is taken. This lesson also holds when we focus on the core RFT subsector of the 5d EFT, as well as on its UV description (UVRT). It was found in \cite{Marchesano:2024tod} that, if the
 tension of the strings in the core RFT lie above $m_{\rm KK,6}^2$, then the UVRT is a 6d SCFT, just like the cases with vertical divisors discussed above. Contrarily, if their tensions lie below the 6d KK scale it is more natural to think of the UVRT as a 5d SCFT.\footnote{The conjecture proposed in \cite{Jefferson:2018irk} predicts that one may understand this 5d SCFT as a 6d SCFT compactified on a circle, possibly with an automorphism twist. It would be interesting to understand how this connection reflects on the rigid curvatures.} 

\subsubsection*{Fibral divisors}

As depicted in figure \ref{fig:tensions}, when $\ker {\bf K}$ contains one or several fibral divisors, the tensions of the core RFT strings lie below  $m_{\rm KK,6}^2$, and so the UVRT is understood as a 5d SCFT. From the viewpoint of the scalar curvature,  when applying the expressions for the divergent pieces obtained in section \eqref{s:rigid}, one finds a geometric picture  very similar to the one obtained in $w=3$ limits. Indeed, let us assume a $w=2$ limit where $\ker {\bf K}$ only contains a fibral divisor, and with a growth sector such that $\gamma_i =\gamma$. We then have the following scalings for the K\"ahler moduli
\be
M^0 \sim \phi^{\frac{\eta}{3}} \,  , \qquad M^E, M^I, M^\mu \sim \phi^{-\frac{2\eta}{3}} \, , 
\label{scalingsMw2}
\ee
and one can see that \eqref{Rdivr1} reduces to
\be
R_{\rm div} = \phi^\eta\,  \kappa_\mu^{-2}\, b^{IJ}\left(\cK_{\mu \mu I} \cK_{\mu \mu J} - \cK_{\mu \mu \mu} \cK_{\mu IJ}\right) + \dots
\label{Rdivr1w2}
\ee
where $\eta = 1 -\gamma$, $\kappa_\mu \simeq (1/\eta_{\bm{e}})^\frac{1}{2} (\cK_{\mu\mu\mu} + \sum_I\cK_{\mu\mu I})$ and $b^{IJ}$ is minus the inverse of $\cK_{eIJ}$. Because in this case we also have that $\cK_{\mu \mu \mu} < 0$, the geometric interpretation is very similar to the one in \eqref{Rdivr1w3}. We will have a positive divergence whenever some divisor in the extended RFT sector $\bm{v}_I$ intersects non-trivially the fibral divisor. Similarly, for $w=2$ limits with a two-dimensional kernel, one can interpret the condition that $\det {\bf M} \neq 0$ as the fact that the intersection of core RFT divisors \eqref{interD} give rise to three independent curves. In this case, from \eqref{Rdivr2} one obtains a divergence of the form $R_{\rm div} \sim \phi^{2\eta}$, while we obtain the subleading divergence $R_{\rm div} \sim \phi^{\eta}$ when $\det {\bf M} = 0$. The main difference with the $w=3$ case is the degree of the divergence, which reflects that the gauge interactions that correspond to the $\bm{v}_I$ sector now go to weak coupling along the limit, instead of remaining constant. Again, the degree of divergence can be expressed in a parametrisation-independent way by expressing it in terms of the quotient $\Lambda_{\rm wgc}/\Lambda_{\rm sp}$. Considering the same simple growth sector $\gamma_i = \gamma$ as above, one finds that 
\be
\Lambda_{\rm sp} \sim \phi^{\frac{\gamma-1}{6}} M_{\rm P}\, ,  \quad \Lambda_{\rm wgc} \sim \phi^{\frac{1-\gamma}{3}} M_{\rm P} \implies \frac{\Lambda_{\rm wgc}}{\Lambda_{\rm sp}} \sim \phi^{\frac{\eta}{2}} \implies R_{\rm div} \sim \left(\frac{\Lambda_{\rm wgc}}{\Lambda_{\rm sp}}\right)^{2\nu} \, ,
\label{scalesw2}
\ee
where $\nu =2$ when we have a leading divergence and $\nu=1$ for a subleading one. Physically, the difference between these two cases is that $\nu=1$ indicates that the core RFT string tension depends on the mass parameters that couple to the SCFT, while $\nu=2$ implies that the gauge kinetic matrix also depends on such mass parameters.

\subsubsection*{Exceptional divisors}

Let us finally consider the case where $\ker {\bf K}$ contains exceptional divisors. As mentioned earlier, in this case the tension of the strings charged under the core RFT sector may lie above or below $m_{\rm KK,6}^2$. For the growth sectors where they lie above, one may consider a circle reduction of the 6d EFT obtained from F-theory on $X$ to compute the 5d EFT moduli space metric, and in particular the curvature of the core RFT subsector. However, unlike for the case of vertical divisors, on can no longer argue that the curvature is non-singular. Indeed, by looking at \eqref{Fgaugecoup} one realises that when an exceptional divisor belongs to $\ker {\bf K}$ the corresponding 6d gauge coupling goes to infinity along the finite-distance 6d trajectory, and we end up in a strong coupling singularity. In fact, this  suggests that the 5d curvature may diverge along the $w=2$ trajectory \eqref{Mtrajw2}. In the following we will confirm this expectation, by using the expressions for the curvature divergences obtained in section \ref{s:rigid}.

As pointed out in \cite{Marchesano:2024tod} for an exceptional divisor to belong to $\ker {\bf K}$ there must be a vertical divisor that does as well, which projects down to the same curve in the base of the fibration. This means that in this case the core RFT that is at least two-dimensional. One may for instance consider as core RFT the toy $su(2)$ model discussed in section 5.2 of  \cite{Marchesano:2024tod}, with prepotential
\be
\cF_{\rm rigid} = -\oh M^E (M^1)^2 + \oh \cK_{E22}M^E (M^2)^2 - M^1 (M^2)^2  + \frac{1}{6} \cK_{222} (M^2)^3\, ,
\ee
where $\cD_E$ corresponds to the zero section and $\cD_\mu$, $\mu=1,2$ belong to the core RFT: $\cD_1$ represents a vertical divisor and $\cD_2$ an exceptional divisor. Applying the expression \eqref{Rdivr2} one can see that   
\be
\det {\bf M} = \det 
\begin{pmatrix}
    0 & 0& \cK_{11}   \\
     0 & -2 & \cK_{12} \\
     -2 & \cK_{222} & \cK_{22} 
\end{pmatrix} 
= - 4 \cK_{11} =  4 M^E \, ,
\ee
while $\det (\cK_{\mu\nu}) = M^E(2M^1-\cK_{E22}M^E-\CK_{222}M^2)- 4(M^2)^2$. In other words, we have that the rigid curvature does not vanish, leading to a divergence along the limit. If we consider the simple growth sector $\gamma_i = \gamma$  that leads to the scalings \eqref{scalingsMw2} we again recover \eqref{scalesw2} with $\nu=2$.  

In this case one should be able to provide a description of the curvature divergence in terms of the physics of 6d SCFTs. By looking at the structure of triple intersection numbers in elliptic fibrations with exceptional divisors, one finds a prepotential for the core RFT of the form \cite{Grimm:2011fx,Bonetti:2011mw}
\be\nonumber
{\cal F}_{\rm rigid}^{\rm cl} =  \oh M^{E} c_{\mu\nu} M^\mu M^\nu + \oh \cK_{E\a\b} M^{E} M^\a M^\b - \oh c_{\mu\mu} C_{\a\b} M^\mu M^\a M^\b + \frac{1}{6} {\cal K}_{\a\b\g} M^\a M^\b M^\b  \, ,
\ee
where the K\"ahler moduli $M^\mu$ correspond to vertical divisors and $M^\a$ to exceptional ones. The tensor $c_{\mu\nu}$ describes the 6d tensor branch metric, while $C_{\a\b}$ is the coroot intersection matrix of the Lie algebra $\mathfrak{g}$ hosted by the exceptional divisors. In the case at hand, we have a rank 2 core RFT with one vertical and one exceptional divisor and $\det {\bf M} \neq 0$ is equivalent to demand that the coroot intersection matrix of the 6d SCFT gauge group is non-trivial. In other words, at least for the case where $\dim \ker {\bf K} =2$, the curvature divergence is associated to the fact that the UVRT is 6d SCFT with a non-Abelian gauge group, which in this case is $su(2)$. This is in agreement with the results of \cite{Marchesano:2024tod}, where the source of curvature divergence in the analogous type IIA limits was identified with the $W$-bosons of the non-Abelian gauge group.  

As done in \cite{Marchesano:2024tod}, by considering more involved growth sectors one may engineer limits where the exceptional divisor is parametrically smaller than the vertical one and the same holds for the tensions of the strings obtained by wrapping such divisors, which can be taken below $m_{\rm KK,6}^2$. Because of this, it is more natural to interpret the RFT sector sourcing the divergence  as a rank-one 5d SCFT associated to the exceptional divisor. Since the rank is one, a maximal divergence cannot be generated and one still gets $\nu=2$. Now the curvature divergence can be interpreted in terms of the cubic CS couplings of this 5d SCFT, which couple it at the level of vevs to the vector multiplets associated to $\cD_1$ and $\cD_E$. As a result, the tension of the string associated to $\cD_2$ depends on some mass parameters. Geometrically, this means that the exceptional divisor is intersected by other effective divisors to generate two effective curve classes, namely the blowup curve in the fibre and the base curve. As for $w=3$ limits, one again expects that their combinations give rise to a cone of effective curves, and then to  towers of bound states of BPS particles becoming massless along the limit. The elements of such towers are charged under the 6d KK $U(1)$, and so we again expect to recover an imbalance between electric and magnetic charges, in the sense specified in section \ref{s:w3}.

\subsubsection*{Summary}

To sum up we find that, for $w=2$ limits, rigid vertical divisors do not give rise to curvature divergences by themselves. They can do so when combined with exceptional divisors, or if replaced by fibral divisors. In the last two cases, which we have analysed for core RFTs of rank $r \leq 2$, the  curvature divergence has a geometrical origin that is very similar to finite-distance 5d limits, which suggests that the it is also linked to the physics of 5d SCFTs. We have that a set of divisors contract to zero size along the limit, and describe a strongly coupled sector under which tensionless strings and infinite towers of massless particles are charged. Whenever the curvature diverges we find that in this massless spectrum there is an imbalance between electric and magnetic charges, similarly to section \ref{s:w3}. Moreover, the degree of the divergence is related to how the core RFT couples to the rest of the vector multiplets through the vevs of their scalars. In particular, if the gauge kinetic matrix depends non-trivially on the non-dynamical mass parameters we have a maximal divergence, and if this is only true for the tension of the core RFT strings we have a subleading one. Notice that all these criteria are not valid for core RFTs made up of only vertical divisors, because in this case their UVRT is always a 6d SCFT. In this case, the curvature divergence is related to the presence of a non-Abelian gauge group, which would require the additional presence of exceptional divisors.

\subsection{Emergent string limits}
\label{ss:w1}

Let us finally consider $w=1$ limits. Depending on the growth sector considered, these may either be emergent strings limits in 5d or a nested limit which first undergoes a decompactification to 6d and then an emergent string limit. It follows from the results of Appendix \ref{ap:asymptotic} that for EFT string limits a divergence for the curvature is linear in the parameter $\phi$, while it goes like $\phi^\eta$, $\eta =1-\gamma$  for the simple growth sector $\gamma_i = \gamma$. Since in this case we can identify the species scale with the square root of the emergent critical string tension $\cT \sim \phi^{-2\eta/3} M_{\rm P}^2$, we find that
\be
\Lambda_{\rm sp} \sim \phi^{-\frac{\eta}{3}} M_{\rm P}\, ,  \quad \Lambda_{\rm wgc} \sim \phi^{\frac{\eta}{6}} M_{\rm P} \implies \frac{\Lambda_{\rm wgc}}{\Lambda_{\rm sp}} \sim \phi^{\frac{\eta}{2}} \implies R_{\rm div} \sim \left(\frac{\Lambda_{\rm wgc}}{\Lambda_{\rm sp}}\right)^{2\nu} \, ,
\label{scalesw1}
\ee
with $\nu=1$. In this case the divergence is at most linear in the parameter $\phi$, which relfects the fact that the extended RFT and core RFT sectors coincide and so do \eqref{Rsublead} and \eqref{Rlead}. This in particular means that a necessary condition to generate a scalar curvature divergence is  $\dim \ker {\bf K} = \dim \ker' {\bf K} +1 \geq 3$. In particular, the RFT building blocks considered in \cite{Marchesano:2024tod} based on simple type II Kulikov degenerations are all free of divergences, since $\dim \ker {\bf K} =2$ for them and so \eqref{Rlead} vanishes identically.

Increasing $\dim \ker {\bf K}$ could in principle remove this feature, although it is easy to see that for some classes of emergent string limits it will still vanish. In particular, for K3 fibrations with only Type II.a Kulikov degenerations, one has that the core RFT can be obtained from dimensional reduction of a 6d tensor branch \cite{Marchesano:2024tod}. Therefore, by the same arguments used in the case of vertical divisors in $w=2$ limits, one expects a core RFT with vanishing curvature. Indeed, in this kind of degenerations the K3 fibre splits into a union of several effective surfaces $S \to \cup_M S_M$, which contain the core RFT sector. Moreover, due to the kind of degeneration they only intersect over the generic elliptic fibre of the Calabi--Yau. This means in particular 
\be
S_N \cdot S_M \cdot S_L = 0\, ,
\ee
which implies that \eqref{Rlead} vanishes identically, independently of the core RFT rank. 
It would then seem that type II.b and type III degenerations with $\dim \ker {\bf K} \geq 3$ are a more suitable setup to build core RFTs with a non-trivial curvature, a task that we leave for the future.


\section{Examples}
\label{s:examples}

We have tested the results discussed above in a series of explicit examples. In this section we present the most interesting cases in some detail, pointing out how they are related to the main discussion. For each of them, after briefly describing the Calabi--Yau geometry, we identify the core and the extended RFT sectors in each limit, we study the asymptotic behaviour of the moduli space curvature and finally discuss the tensions and charge-to-mass ratios of the BPS objects charged under the core RFT.

\subsection{KMV conifold}

The first example that we discuss is the KMV conifold \cite{Klemm:1996hh,Alim:2021vhs}. The toric data and the GLSM matrix for this model are given by
\setlength{\arrayrulewidth}{0.2mm}
\renewcommand{\arraystretch}{0.8}
\begin{table}[H]
\begin{center}
\begin{tabular}{cccc|ccc}
& & & & ${\cal C}^1$ & ${\cal C}^2$ & ${\cal C}^3$ \\
\hline
1 & 0 & 0 & 0 & 0 & 0 & 1 \\
$-1$ & $-1$ & $-6$ & $-9$ & 0 & 0 & 1 \\
0 & 1 & 0 & 0 & 0 & 1 & 0 \\
0 & $-1$ & $-4$ & $-6$ & 0 & 1 & $-1$ \\
0 & 0 & 1 & 0 & 2 & 0 & 0 \\
0 & 0 & 0 & 1 & 3 & 0 & 0 \\
0 & 0 & $-2$ & $-3$ & 1 & $-2$ & $-1$
\end{tabular}
\end{center}
\label{tab:toric,GLSM,KMV}
\end{table}
\noindent This manifold has two different topological phases, that we call $\mathbb{F}_1$ and $\mathbb{P}^2$ phases, as anticipated in Section \ref{s:w3}. We start with the $\mathbb{F}_1$ phase, which is a smooth elliptic fibration with base $\mathbb{F}_1$. Expanding the K\"ahler form in a basis of K\"ahler cone generators $J=M^a J_a$, $a=1,2,3$, the moduli $\{M^a\}$ are the volumes of the Mori cone generators $\{{\cal C}^a\}$ in 11-dimensional Planck units and we have the following triple intersection polynomial
\be
\cI(X) = 8 J_1^3 + J_1^2 \left( 3J_2 + 2J_3 \right) + J_1 \left( J_2^2 + J_2 J_3 \right) \, .
\ee
In this basis $M^1=M^E$ parametrises the volume of the elliptic fibre, while $\{M^2,M^3\}$ parametrise the volumes of the fibre $\mathbb{P}^1_f$ and base $\mathbb{P}^1_b$ of $\mathbb{F}_1$, respectively. We first consider a finite distance trajectory of the form \eqref{Mtrajw3} with $e_0^a = \del^a_1$, which geometrically corresponds to keeping the elliptic fibre with finite constant area while shrinking the base $\mathbb{F}_1$ to a point, reaching a strong coupling singularity with a rank $r=1$ 5d SCFT. For this limit we have
\be
\bK_{ab} =
\begin{pmatrix}
    8 & 3 & 2 \\
    3 & 1 & 1 \\
    2 & 1 & 0
\end{pmatrix} \, ,
\qquad {\bf k}_a = (8,3,2) \, , \qquad {\bf k} = 8 \, .
\ee
We then perform the change of basis to  \eqref{basisw3}, given by
\be     \label{basisKMV}
\cD_0 = J_1 \, , \qquad \cD_I = 2J_2-3J_3 \, , \qquad \cD_\m = -J_1+2J_2+J_3 \, ,
\ee
where $\cD_\mu \in \ker {\bf K}$ corresponds to (minus) the collapsing  base divisor $\mathbb{F}_1$. Notice that, given the definitions \eqref{basisw3}, we always have the freedom to shift $\cD_I$ by a multiple of the kernel vector $\cD_\m$, and that we have used this freedom to pick $\cD_I$ such that $\cK_{\m\m I}=0$, as in \eqref{DIprime}. The core RFT sector is given by the kernel $\cD_\m$ and the extended RFT also includes the direction $\cD_I$, which is a non-effective divisor class. One can check that for these directions the rigid limit condition \eqref{rigid} is satisfied. The matrix $\cK^{ab}$ in this basis reads
\be
\cK^{ab} \simeq
\begin{pmatrix}
    \frac{1}{8 M^0} & -\frac{M^I}{8 (M^0)^2} & \frac{(M^I)^2}{8 M^\m (M^0)^2} \\
    -\frac{M^I}{8 (M^0)^2} & - \frac{1}{8 M^0} & \frac{M^I}{8 M^\m M^0} \\
    \frac{(M^I)^2}{8 M^\m (M^0)^2} & \frac{M^I}{8 M^\m M^0} & -\frac{1}{8 M^\m}
\end{pmatrix} 
\sim
\begin{pmatrix}
    \text{const} & \phi^{-1} & \phi^{-1} \\
    \phi^{-1} & \text{const} & \text{const} \\
    \phi^{-1} & \text{const} & \phi \\
\end{pmatrix}\, ,
\ee
in agreement with the analysis of Appendix \ref{ap:asymptotic} for $w=3$ EFT string limits. It easy to see that a would-be leading divergence in the moduli space curvature \eqref{RMth2} scaling as $\phi^3$ should come from a cubic term in the kernel component $\cK^{\m\m}$. However, since there is only one kernel element, this contribution is cancelled by the antisymmetrisation of indices, and we have at most a quadratic divergence. Indeed, the moduli space curvature asymptotically goes like
\be
R_{\rm M} \simeq \frac{1}{8 (M^\m)^2 M^0} \sim \phi^2 \, ,
\ee
which coincides with the expressions in \eqref{Rdivr1} and \eqref{Rdivr1w3}, with
\be
\hat{\cK}^{IJ} \simeq -\frac{1}{8 M^0}\, , \qquad \cK_{\m\m} \simeq -8M^\m \, , \qquad \cK_{\m\m\m} = -\cK_{\m II} = \cK_{0II} = -8 \, ,\qquad \cK_{\m\m I} = 0 \, .
\ee
In particular, the necessary condition for a curvature divergence that the core RFT divisors intersect some divisors in the extended RFT is satisfied, since $\cK_{\m II} \neq 0$.

The BPS strings that are part of the core RFT are those arising from M5-branes wrapping the base $\mathbb{F}_1$, for which we have $\bm{p} = (1,-2,-1)$ in the original basis $\{J_a\}$ and
\be
T_{\bm{p}}/M_{\rm P}^2 \propto (M^\m)^2 - (M^I)^2 \, , \qquad \g_{\bm{p}}^2 \simeq \frac{M^\m}{\left( (M^\m)^2-(M^I)^2 \right)^2} \sim \phi^3 \, ,
\ee
along the trajectory we are considering. This charge direction is the only one for which the charge-to-tension ratio diverges cubically. In other words, as soon as we consider a string that is not only charged under the core RFT the divergence is suppressed. This is not true for particles, since for any particle with charges $\bm{q} = (0,q_2,q_3)$, $q_2,q_3 \geq 0$ we have
\be
\g_{\bm{q}}^2 \simeq \frac{\left( 2q_2+q_3 \right)^2}{4 M^\m \left( 2q_2 (M^\m + M^I) + q_3 (M^\m -3 M^I) \right)^2} \sim \phi^3 \, .
\ee
These particles correspond to M2-branes wrapping effective curves inside the base $\mathbb{F}_1$ and even if not all these curves are actually populated by BPS states (see figure \ref{fig:GV_KMV}), they span a 2-dimensional cone of charges. The particles that are only charged under the core RFT $U(1)$ correspond to the charge direction $\bm{q}=(0,3,2)$, while all the rest are charged under both the $U(1)$'s of the extended RFT. Since the cubic divergence in the curvature is absent, while there exist BPS states with $\g^2 \sim \phi^3$, the relation \eqref{Rgmax} is satisfied. Notice also that the tension and mass of these BPS objects depends on the mass parameter $M^I$. The core RFT gauge kinetic function, on the other hand, does not, as expected in absence of maximal divergences in the curvature.

As anticipated, this geometry admits a flop transition to a different phase $\tilde{X}$, that we call $\mathbb{P}^2$ phase, realised by shrinking to zero volume the curve ${\cal C}^3 = \mathbb{P}^1_b$ and by blowing up a different $\mathbb{P}^1$. The new Mori cone generators in the flopped phase are $\{\tilde{\cal C}^a\} = \{{\cal C}^1+{\cal C}^3,{\cal C}^2+{\cal C}^3, -{\cal C}^3\}$ and the triple intersection polynomial, in the dual basis of K\"ahler cone generators $\{\tilde{J}_a\}$, reads
\be
\cI(\tilde{X}) = 8 \tilde{J}_1^3 + \tilde{J}_1^2 \left( 3\tilde{J}_2 + 9\tilde{J}_3 \right) + \tilde{J}_1 \left( \tilde{J}_2^2 + 3\tilde{J}_2 \tilde{J}_3 + 9\tilde{J}_3^2 \right) + \tilde{J}_2^2 \tilde{J}_3 + 3 \tilde{J}_2 \tilde{J}_3^2 + 9 \tilde{J}_3^3 \, .
\ee
In this phase the base $\mathbb{F}_1$ becomes a $\mathbb{P}^2$ surface, whose volume is controlled by $\tilde{M}^2$. Furthermore the elliptic fibration becomes non-flat, because the dimension of the fibre jumps over points of the base. The fibral divisor over these points is a del Pezzo surface $dP_8$, whose volume is controlled by the modulus $\tilde{M}^1$. Let us consider, for instance, the finite distance limit in which $\tilde{M}^3$ is sent to a constant, while $\tilde{M}^1,\tilde{M}^2$ go to zero. In this limit both the base $\mathbb{P}^2$ and the fibral divisor $dP_8$ are shrinking to zero volume and we reach a rank $r=2$ 5d SCFT point. By taking the following basis of divisors
\be
\cD_0 = \tilde{J}_3 \, , \qquad \cD_\m = 3\tilde{J}_2-\tilde{J}_3 \, , \qquad \cD_\n = \tilde{J}_1 - \tilde{J}_3  \, ,
\ee
where $\cD_\m = -\mathbb{P}^2$ and $\cD_\n = -dP_8$, the triple intersection numbers and $\cK^{ab}$ diagonalise
\be     \label{Kabc_diag}
\cK_{000} = 9 \, , \qquad \cK_{\m\m\m} = -1 \, , \qquad \cK_{\n\n\n} = -9 \, ,
\ee
\be     \label{Kab_diag}
\cK^{ab} = 
\begin{pmatrix}
 \frac{1}{9 M^0} & 0 & 0\\
 0 & -\frac{1}{M^\m} & 0 \\
 0 & 0 & -\frac{1}{9 M^\n}
\end{pmatrix}
\sim
\begin{pmatrix}
    \text{const} & 0 & 0 \\
    0 & \phi & 0 \\
    0 & 0 & \phi \\
\end{pmatrix}\, ,
\ee
which implies that there cannot be any curvature divergence, due to the antisymmetrisation of indices  in \eqref{RMth2}. There are nonetheless BPS objects charged under the RFT with divergent charge-to-mass ratios. In particular, for strings with charges $\bm{p}_\m = (0,-3,1)$ and $\bm{p}_\n = (-1,0,1)$, i.e. M5's wrapping $\mathbb{P}^2$ and $dP_8$, we have
\be
\begin{split}
& T_{\bm{p}_\m}/M_{\rm P}^2 \propto (M^\m)^2 \, , \qquad \g_{\bm{p}_\m}^2 \simeq \frac{8}{9\left( M^\m \right)^3} \sim \phi^3 \, ,\\
& T_{\bm{p}_\n}/M_{\rm P}^2 \propto (M^\n)^2 \, , \qquad \g_{\bm{p}_\n}^2 \simeq \frac{8}{\left( M^\n \right)^3} \sim \phi^3 \, .
\end{split}
\ee
As regards BPS particles, any M2-brane wrapping a combination of $\tilde{\cal C}^1$ and $\tilde{\cal C}^3$ would have a cubically divergent charge-to-mass ratio. However, looking at the GV invariants \cite{Alim:2021vhs}, we see that the only curves that actually host BPS states are the multiwrappings of either $\tilde{\cal C}^1$ or $\tilde{\cal C}^3$, which are proportional to the self-intersection of $\mathbb{P}^2$ and $dP_8$, respectively. So the BPS particles with divergent $\g$ do not really span a 2-dimensional cone of charges, but rather two 1-dimensional rays. We can also interpret this in the following way. The rank-2 SCFT that we reach at the end of the trajectory can be seen as two rank-1 SCFTs that are fully decoupled from one another, as it is clear from \eqref{Kabc_diag} and \eqref{Kab_diag}, as well as from any other vector multiplet. Due to this decoupling, no curvature divergence can be generated and \eqref{Rgmax} is satisfied trivially.


\subsection{Genus-one fibration with exceptional divisors}

We now turn to a more involved example that illustrates several features of infinite distance limits. We consider a Calabi--Yau threefold with $h^{(1,1)}=5$ studied in \cite{Lee:2018spm}, which can be viewed as an elliptic fibration over $\mathbb{F}_1$ with two extra sections. In a basis of Mori cone generators $\{{\cal C}^a\}$, we have that ${\cal C}^2=\mathbb{P}^1_b$ and ${\cal C}^4=\mathbb{P}^1_f$ are the base and fibre of the $\mathbb{F}_1$ base, while ${\cal C}^1,{\cal C}^3,{\cal C}^5$ are fibral curves. The toric data and the GLSM matrix for this geometry are
\setlength{\arrayrulewidth}{0.2mm}
\renewcommand{\arraystretch}{0.8}
\begin{table}[H]
\begin{center}
\begin{tabular}{cccc|ccccc}
  &  &  &  & ${\cal C}^1$ & ${\cal C}^2$ & ${\cal C}^3$ & ${\cal C}^4$ & ${\cal C}^5$ \\
\hline
$-1$ & 1 & 0 & 0 & 1 & $-1$ & $-1$ & 0 & 1 \\
0 & $-1$ & 0 & 0 & 0 & 0 & 0 & 0 & 1\\
1 & 0 & 0 & 0 & 1 & 0 & 0 & 0 & 0\\
$-1$ & 0 & 0 & 0 & 0 & 0 & 1 & $-2$ & $-1$\\
0 & 1 & 0 & 0 & $-1$ & 0 & 1 & 0 & 0\\
$-1$ & 1 & 1 & 0 & 0 & 1 & 0 & 0 & 0\\
0 & 0 & $-1$ & $-1$ & 0 & 1 & 0 & 0 & 0\\
$-2$ & 0 & 0 & 1 & 0 & 0 & 0 & 1 & 0 \\
0 & 0 & 0 & $-1$ & 0 & $-1$ & 0 & 1 & 0 \\
\end{tabular}
\end{center}
\label{tab:toric,GLSM,5mod}
\end{table}
\noindent In the basis of K\"ahler cone generators $\{J_a\}$ dual to $\{\mathcal{C}^a\}$, the intersection polynomial reads
\be
\begin{split}
\mathcal{I}(X) = & 16 J_1^3 + 4 J_1^2 J_2 + 30 J_1^2 J_3 + 6 J_1^2 J_4 + 18 J_1^2 J_5 + 8 J_1 J_2 J_3 + 2 J_1 J_2 J_4 + 4 J_1 J_2 J_5 \\& +40 J_1 J_3^2 +
+ 11 J_1 J_3 J_4 + 18 J_1 J_3 J_5 + 2 J_1 J_4^2 + 7 J_1 J_4 J_5 + 4 J_1 J_5^2 + 10 J_2 J_3^2 \\
& +3 J_2 J_3 J_4 + 4 J_2 J_3 J_5 + 2 J_2 J_4 J_5 + 44 J_3^3 + 13 J_3^2 J_4 + 18 J_3^2 J_5 + 3 J_3 J_4^2  \\ & + 7 J_3 J_4 J_5 + 4 J_3 J_5^2 + 2 J_4^2 J_5 + 2 J_4 J_5^2 \, .
\end{split}
\ee
We consider the $w=2$ limit given by $e_0^a = \del_4^a$, which geometrically corresponds to taking the fibre of the base ${\cal C}^4$ to infinite volume, while shrinking the other curves to zero size in 5-dimensional Planck units. In this limit we have
\be
\bK_{ab} =
\begin{pmatrix}
    6 & 2 & 11 & 2 & 7 \\
    2 & 0 & 3 & 0 & 2 \\
    11 & 3 & 13 & 3 & 7 \\
    2 & 0 & 3 & 0 & 2 \\
    7 & 2 & 7 & 2 & 2
\end{pmatrix} \, ,
\qquad {\bf k}_a = (2,0,3,0,2) \, ,
\ee
and we then perform a change of basis as in \eqref{basisw2}, which in this case gives
\be     \label{basis5mod}
\begin{split}
\cD_E = 2J_3-J_4&-2J_5 \, , \qquad \cD_0 = J_4 \, , \qquad \cD_I = J_1-2J_2-2J_3+2J_5 \, ,\\
&\cD_J = J_5-J_1-J_2 \, , \qquad \cD_\m = J_2-J_4 \, .
\end{split}
\ee
The core RFT is given by the kernel divisor $\cD_\m$, which is (minus) a vertical divisor over the curve $\mathbb{P}^1_b$, while the extended RFT also includes the field directions associated to $\cD_I,\cD_J$, which are the images under the Shioda map of two extra rational sections. Since the kernel only contains a vertical divisor, by the discussion of section \ref{ss:w2} we do not expect any curvature divergence, and indeed in this limit we have
\be
R_{\rm M} \simeq -\frac{7878}{1225} \, .
\ee
Even though we do not have curvature divergences, we still have BPS objects with divergent charge-to-mass ratios. The leading divergences are obtained for the string with charges $\bm{p}=(0,-1,0,1,0)$ and its multiples
\be
\begin{split}
& T_{\bm{p}}/M_{\rm P}^2 \propto 2 M^E (M^\m - M^E) - (M^I)^2 - M^I M^J - (M^J)^2 \, ,\\
& \g_{\bm{p}}^2 \simeq \frac{4 M^E}{\left( 2 M^E (M^\m - M^E) - (M^I)^2 - M^I M^J - (M^J)^2 \right)^2} \sim \phi^2 \, ,
\end{split}
\ee
so \eqref{Rgmax} is trivially satisfied. Furthermore, the gauge kinetic matrix of the core RFT reads
\be
\cK_{\m\m} \simeq -2 M^E \, .
\ee
Notice that, from the viewpoint of a 5d SCFT, we would have a tension and gauge kinetic term that depend on mass parameters. However, as discussed in section \ref{ss:w2} the UVRT of this core RFT should be identified with a 6d SCFT. One should then apply a different criterion, namely the presence of a non-Abelian gauge group, which is absent in this case.

Let us now take a different $w=2$ limit, given by $e_0^a = \del^a_5$, where we grow the curve ${\cal C}^5$. In order to understand this as an F-theory limit, one should notice that, in addition to the fibration structure described above, this model also admits an alternative, incompatible, genus-one fibration over $\mathbb{F}_1$ with two exceptional divisors and one bisection. In this second description, the base $\mathbb{F}_1$ contains the curves ${\cal C}^2=\mathbb{P}^1_b$, as before, and ${\cal C}^5=\mathbb{P}^1_f$, while ${\cal C}^1,{\cal C}^3,{\cal C}^4$ are the fibral curves that make up the genus-one fibre. Hence growing the curve ${\cal C}^5$ takes us to an F-theory regime. In this limit we have
\be
\bK_{ab} =
\begin{pmatrix}
    18 & 4 & 18 & 7 & 4 \\
    4 & 0 & 4 & 2 & 0 \\
    18 & 4 & 18 & 7 & 4 \\
    7 & 2 & 7 & 2 & 2 \\
    4 & 0 & 4 & 2 & 0
\end{pmatrix} \, ,
\qquad {\bf k}_a = (4,0,4,2,0) \, ,
\ee
and we pick the following basis
\be
\begin{split}
\cD_E = J_5 +2J_4&-2J_2 \, , \qquad \cD_0 = J_5 \, , \qquad \cD_I = 2J_3-4J_4-3J_5 \, ,\\
&\cD_\m = J_2-J_5 \, , \qquad \cD_\n = J_2+J_3-J_1-J_5 \, .
\end{split}
\ee
The core RFT  has rank 2 and contains the field directions associated to the kernel divisors $\cD_\m,\cD_\n$, that are (minus) a vertical and an exceptional over $\mathbb{P}^1_b$, respectively. Since there is an exceptional divisor in the kernel, from the discussion in section \ref{ss:w2} we expect a  curvature divergence. In particular, given the following triple intersections among its elements
\be
\cK_{\m\n\n} = -2 \, , \qquad \cK_{\n\n\n} = -8 \, ,
\ee
we expect the divergence to have the maximum degree for this class of limits and indeed one can check that the curvature asymptotically scales like
\be
R_{\rm M} \simeq \frac{M^E}{\left( (M^E)^2 + 2 M^E (M^\m + 3 M^\n) - (M^\n)^2 \right)^2} \sim \phi^2\, .
\ee
Furthermore, as the exceptional divisor shrinks to zero size, we recover a non-abelian gauge group, which in this case is $SU(2)$, whose W-boson is given by an M2-brane wrapping the fibral curve contained in the exceptional divisor. The strings obtained from M5-branes wrapping the kernel divisors have the following tensions and ratios:
\be
\begin{split}
& T_{\bm{p}_\m}/M_{\rm P}^2 \propto (M^\n)^2 + 2 M^E (2 M^\m + M^\n) - 4 (M^E)^2 \, ,\\
& \g_{\bm{p}_\m}^2 \simeq \frac{8 M^E}{\left( (M^\n)^2 + 2 M^E (2 M^\m + M^\n) - 4 (M^E)^2 \right)^2} \sim \phi^2 \, ,
\end{split}
\ee
\be
\begin{split}
& T_{\bm{p}_\n}/M_{\rm P}^2 \propto (M^\m + 2 M^\n - M^E) (M^E + M^\n) \, ,\\
& \g_{\bm{p}_\n}^2 \simeq \frac{M^E + M^\m +4 M^\n}{\left( (M^\m + 2 M^\n - M^E) (M^E + M^\n) \right)^2} \sim \phi^2 \, ,
\end{split}
\ee
from where we see that \eqref{Rgmax} is satisfied. The gauge kinetic matrix of the core RFT reads
\be
\cK_{\m\n} = 
\begin{pmatrix}
    -4 M^E & -2 (M^E + M^\n) \\
 -2 (M^E + M^\n) & -2 (M^E + M^\m +4 M^\n)
\end{pmatrix} \, .
\ee

This geometry also admits a flop transition to a different phase, realised by shrinking the curve ${\cal C}^2 = \mathbb{P}^1_b$ to zero and blowing up a different $\mathbb{P}^1$. In the new phase we no longer have a flat genus-one fibration, similarly to the flopped phase of the KMV conifold considered before. The Mori cone generators of this phase are $\{\tilde{\cal C}^a\} = \{{\cal C}^1+{\cal C}^2, - {\cal C}^2,{\cal C}^3, {\cal C}^4+{\cal C}^2, {\cal C}^5 + {\cal C}^2\}$ and in the dual basis of K\"ahler cone generators the triple intersection numbers are given by
\be
\begin{split}
\mathcal{I}(\tilde{X}) = & 16 \tilde{J}_1^3 + 36 \tilde{J}_1^2 \tilde{J}_2 + 30 \tilde{J}_1^2 \tilde{J}_3 + 6 \tilde{J}_1^2 \tilde{J}_4 + 18 \tilde{J}_1^2 \tilde{J}_5 + 64 \tilde{J}_1 \tilde{J}_2^2 + 51 \tilde{J}_1 \tilde{J}_2 \tilde{J}_3 + 13 \tilde{J}_1 \tilde{J}_2 \tilde{J}_4  \\
& + 25 \tilde{J}_1 \tilde{J}_2 \tilde{J}_5 + 40 \tilde{J}_1 \tilde{J}_3^2 + 11 \tilde{J}_1 \tilde{J}_3 \tilde{J}_4 + 18 \tilde{J}_1 \tilde{J}_3 \tilde{J}_5 + 2 \tilde{J}_1 \tilde{J}_4^2 + 7 \tilde{J}_1 \tilde{J}_4 \tilde{J}_5 + 4 \tilde{J}_1 \tilde{J}_5^2 + 101 \tilde{J}_2^3 \\
&+ 79 \tilde{J}_2^2 \tilde{J}_3 + 22 \tilde{J}_2^2 \tilde{J}_4 + 34 \tilde{J}_2^2 \tilde{J}_5 + 61 \tilde{J}_2 \tilde{J}_3^2 + 18 \tilde{J}_2 \tilde{J}_3 \tilde{J}_4 + 25 \tilde{J}_2 \tilde{J}_3 \tilde{J}_5 + 4 \tilde{J}_2 \tilde{J}_4^2 + 9 \tilde{J}_2 \tilde{J}_4 \tilde{J}_5 \\
&+ 6 \tilde{J}_2 \tilde{J}_5^2 + 44 \tilde{J}_3^3 + 13 \tilde{J}_3^2 \tilde{J}_4 + 18 \tilde{J}_3^2 \tilde{J}_5 + 3 \tilde{J}_3 \tilde{J}_4^2 + 7 \tilde{J}_3 \tilde{J}_4 \tilde{J}_5 + 4 \tilde{J}_3 \tilde{J}_5^2 + 2 \tilde{J}_4^2 \tilde{J}_5 + 2 \tilde{J}_4 \tilde{J}_5^2 \, .
\end{split}
\ee
Let us consider the $w=2$ limit in which we grow the curve $\tilde{\cal C}^4$ and shrink all the rest to zero volume, given by the trajectory \eqref{Mtrajw2} with $e^a_0 = \del^a_4$. For this trajectory
\be
\bK_{ab} =
\begin{pmatrix}
    6 & 13 & 11 & 2 & 7 \\
    13 & 22 & 18 & 4 & 9 \\
    11 & 18 & 13 & 3 & 7 \\
    2 & 4 & 3 & 0 & 2 \\
    7 & 9 & 7 & 2 & 2
\end{pmatrix} \, ,
\qquad {\bf k}_a = (2,4,3,0,2) \, ,
\ee
and we pick a similar basis to \eqref{basis5mod}
\be
\begin{split}
\cD_E = -2\tilde{J}_1 +2 \tilde{J}_2 + 2 & \tilde{J}_3 -\tilde{J}_4 - 4\tilde{J}_5 \, , \qquad \cD_0 = \tilde{J}_4 \, , \qquad \cD_I = 4\tilde{J}_2-\tilde{J}_1-2\tilde{J}_3 -2\tilde{J}_4 -2\tilde{J}_5\, ,\\
&\cD_J = \tilde{J}_5-\tilde{J}_1-\tilde{J}_4 \, , \qquad \cD_\m = \tilde{J}_1-\tilde{J}_2 + \tilde{J}_5 \, ,
\end{split}
\ee
where we performed a shift in $\cD_E$, $\cD_I$ and $\cD_J$ with respect to \eqref{basis5mod}, to make them orthogonal to $\cD_\m$ as discussed around \eqref{DIprime}. Since the core RFT has rank 1, the leading divergence is absent, due to the antisymmetrisation of indices in the curvature expression. However, the kernel divisor $\cD_\m$ has a different topology after the flop and is no longer a vertical divisor. In particular, it admits non-vanishing intersections with $\cD_I$ and $\cD_J$
\be
\cK_{\m II} = \cK_{\m JJ} = 2 \, , \qquad \cK_{\m IJ} = -1 \, ,
\ee
which give rise to a subleading divergence. One can check this by computing explicitly the moduli space curvature, that in this limit behaves as
\be
R_{\rm M} \simeq \frac{22}{35 (M^\m)^2 M^0} \sim \phi \, .
\ee
The M5 wrapping the core RFT divisor $\cD_\m$ has the following tension and charge-to-tension ratio
\be
\begin{split}
& T_{\bm{p}_\m}/M_{\rm P}^2 \propto (M^\m)^2 -2 (M^J)^2 + 2 M^J ( 2 M^E + M^I ) - 2 ( 2 M^E + M^I )^2 \, ,\\
& \g_{\bm{p}_\m}^2 \simeq \frac{8 M^\m}{\left( (M^\m)^2 -2 (M^J)^2 + 2 M^J ( 2 M^E + M^I ) - 2 ( 2 M^E + M^I )^2 \right)^2} \sim \phi^2 \, ,
\end{split}
\ee
so the relation \eqref{Rgmax} is satisfied. The gauge kinetic matrix of the RFT reads
\be
\cK_{\m\m} = - M^\m \, ,
\ee
which is independent of the mass parameters, as expected in absence of a quadratic divergence in the curvature.


\section{Conclusions}
\label{s:conclu}

In this work we have analysed the moduli space curvature of M-theory compactified on a Calabi--Yau threefold, focusing on the vector multiplet sector of the resulting 5d $\cN=1$ EFT. We first computed the scalar curvature of such a moduli space, by relating it to the curvature of the saxionic slice of the 4d EFT obtained from type IIA on the same Calabi--Yau. We then studied its behaviour along finite and infinite distance trajectories of the form \eqref{growth}, that correspond to the 5d uplift of EFT string limits in 4d and more general growth sectors.

From the general expression of the curvature, we found that in order to have a divergence one needs some sector of the gauge fields to go to strong coupling and this can only be realised in the presence of a non-trivial kernel for the matrix $\bK$ defined in \eqref{mK} (excluding the limit direction in $w=1$ limits, which cannot give rise to any divergence). The kernel of such matrix determines what we call the \textit{core RFT}, i.e. a gauge theory sector that goes to strong coupling as it reaches a rigid field theory regime, as explained around \eqref{rigid}. Requiring to have a kernel and using it to identify the RFT sector is completely  analogous to the 4d analysis performed in \cite{Marchesano:2023thx}. However, the conditions to actually have a curvature divergence are more subtle in 5d. For the 5d uplift of an EFT string limit with scaling weight $w$ and a non-trivial core RFT, the maximal possible divergence in the curvature is $R_{\rm M} \sim \phi^w$, which corresponds to the  core RFT curvature. In many examples this leading divergence vanishes, simply because the core RFT has a vanishing moduli space curvature. This happens, for instance, if the core RFT has rank one, which implies that its moduli space is one-dimensional, but also in other, more involved cases. We then consider possible subleading divergences, that arise from the curvature of what we call the \textit{extended RFT}, which contains the core RFT plus additional field directions where the rigid field theory relations are recovered asymptotically, but do not go to strong coupling.

Our analysis reveals that the conditions to determine whether the maximal divergence is present in the curvature or not depend on the UV rigid completion (or UVRT) of the core RFT in the limit we are considering. If the UVRT is a 5d SCFT, then a maximal divergence is present when the core RFT gauge kinetic matrix $\cK_{\m\n}$  depends on some non-dynamical mass parameters. If this does not happen, but the tensions of the BPS strings charged under the core RFT do depend on them, then we have a subleading divergence. If instead there is a full decoupling between the core RFT and the rest of vector multiplets at the level of the cubic Chern--Simons terms, then no curvature divergence is present. In the case where the UVRT is a 6d SCFT, in order to have a curvature divergence we instead need a non-Abelian gauge group in the UVRT. Let us mention that we have focused most of our analysis on core RFTs of rank $r \in \{1,2\}$. For higher ranks a general analysis becomes much more involved and we did not find any explicit examples. It would be very interesting to check our results in these more complicated setups, in order to fully understand what UV information is captured by the moduli space curvature. Additionally, it would be interesting to explore how our analysis extends to the hypermultiplet sector of the EFT, whose infinite distance limits should be related to their type II analogues \cite{Marchesano:2019ifh,Grimm:2019wtx,Baume:2019sry}. In particular it would be interesting to check whether regions with classical curvature divergences are obstructed by quantum corrections, along the lines of  \cite{Marchesano:2019ifh,Baume:2019sry}.

The field directions that belong to the core and extended RFT can also be identified by looking at the charge-to-tension ratios $\g_{\bm{p}}$ of BPS strings that are magnetically charged under these sectors. As explained in section \ref{ss:gamma}, such ratios should diverge for objects that are only charged under $U(1)$'s  reaching an RFT regime. Among the divergent $\g_{\bm{p}}$, the highest degree of divergence selects the core RFT sector, while subleading divergences correspond to the extended RFT. The similarities of these results with the findings in \cite{Castellano:2024gwi} for 4d $\cN=2$ theories point towards an overarching picture that applies in very different setups, and that agrees with the Curvature Criterion proposed in \cite{Marchesano:2023thx}, or some generalisation thereof.

\bigskip

\centerline{\bf  Acknowledgments}

\vspace*{.5cm}

We would like to thank Alberto Castellano, Ben Heidenreich, Stefano Lanza, Lorenzo Paoloni, Tom Rudelius and Max Wiesner for discussions.  This work is supported through the grants CEX2020-001007-S and PID2021-123017NB-I00, funded by MCIN/AEI/10.13039/501100011033 and by ERDF A way of making Europe. AB is supported through the JAE Intro grant JAEINT\_23\_01949. LM is supported by the fellowship LCF/BQ/DI21/11860035  from ``La Caixa" Foundation (ID 100010434).


\appendix


\section{Asymptotic expansion of gauge kinetic terms}
\label{ap:asymptotic}

In this appendix we define the field directions that belong to the gravitational sector and to the extended and core RFTs for EFT string limits of scaling weight $w$, uplifted to 5d variables. In addition, we compute the asymptotic scaling of the gauge kinetic matrix $\cI_{ab}$, as well as its rigid version  $\cK_{ab}$, and their inverses. Finally, in each case we prove that the only field directions that can contribute to curvature divergences are those of the extended RFT. Our discussion focuses on EFT string limits, but our findings qualitatively hold also for more general growth sectors.

\subsubsection*{$w=3$ limits}

In this class of limits we pick the basis of divisors $\{\bm{v}_0, \bm{v}_I, \bm{v}_\m\}$ defined by
\be     \label{basisw3}
\bm{v}_0 = \bm{e}_0\, , \qquad v_I^a {\bf k}_a = 0\, , \text{but } \bm{v}_I \notin \ker \bK\, , \qquad \bm{v}_\m \in \ker \bK \, .
\ee
To connect with the discussion in section \ref{ss:decoupling}, $\bm{v}_0$ can be identified with the graviphoton multiplet, $\bm{v}_I$ and $\bm{v}_\m$ make up the extended RFT, while the core RFT only contains the kernel directions $\bm{v}_\m$. Notice that the above definitions imply that
\be
\cK_{0\m a} = \cK_{abc} e_0^b v_\m^c = 0 \, , \qquad  \cK_{00I} = \cK_{abc} e_0^a e_0^b v_I^c = 0 \, ,
\ee
and at leading order the matrix $\cK_{ab}$ and the vector $\cK_a$ take the following form
\be     \label{Kab_w3}
\cK_{ab} \simeq
\begin{pmatrix}
    \cK_{000} M^0 & \cK_{0IJ} M^J & 0 \\
    \cK_{0IJ} M^J & \cK_{0IJ} M^0 & \cK_{I\m a} M^a \\
    0 & \cK_{I\m a} M^a & \cK_{\m\n a} M^a
\end{pmatrix} \sim
\begin{pmatrix}
    \text{const} & \phi^{-1} & 0 \\
    \phi^{-1} & \text{const} & \phi^{-1} \\
    0 & \phi^{-1} & \phi^{-1}
\end{pmatrix}\, ,
\ee
\be
\cK_a \simeq \left( \cK_{000} (M^0)^2 \, , \,  \cK_{0IJ} M^0 M^J \, , \, \cK_{\m ab} M^a M^b \right) \sim \left( \text{const} \, , \phi^{-1} \, , \phi^{-2} \right)\, ,
\ee
where the sums over $a$ do not include the direction of the limit $M^0$ and the leading term of the moduli is given by \eqref{Mtrajw3}, in particular at leading order $M^0 \sim \text{const}$ while all the other moduli are suppressed by $\phi^{-1}$. With this information we can now compute the scaling of the components of the gauge kinetic matrix \eqref{Iabsugra}
\be
\cI_{ab} \sim \begin{pmatrix}
    \text{const} & \phi^{-1} & \phi^{-2} \\
    \phi^{-1} & \text{const} & \phi^{-1} \\
    \phi^{-2} & \phi^{-1} & \phi^{-1}
\end{pmatrix} \, ,
\ee
which can be used to determine the behaviour of the charge-to-tension ratios of BPS strings \eqref{gammastring}, which asymptotically scale like
\be
\g^2_0 \sim \text{const} \, , \qquad \g^2_I \sim \phi^2 \, , \qquad \g^2_\m \sim \phi^3 \, .
\label{apgammaw3}
\ee
If a string is charged under more than one of the sectors above, the overall behaviour of the charge-to-tension ratio is selected by the least divergent direction. Furthermore, one can see that for the directions $\bm{v}_I$ and $\bm{v}_\m$ the gauge kinetic matrix asymptotes to the rigid field theory expression, i.e. $\cI_{mn} \simeq -\cK_{mn}$, and the ratios diverge, as we expect for RFT directions.

In order to determine which sectors contribute to curvature divergences, we use the expression \eqref{RMth2} and we need to compute the inverse of \eqref{Kab_w3}. In order to do that we first compute its determinant, which at leading order reads
\be
\det \cK_{ab} \simeq \cK_{000} \det \left( \cK_{0IJ} \right) \det \left( \cK_{\m\n a} M^a \right) \left(M^0\right)^{n_I+1} \sim \phi^{-r}\, ,
\ee
where $n_I$ and $r$ are the number of independent vectors of type $v_I$ and the dimension of the kernel (or the rank of the core RFT), respectively. Notice that $\det \left( \cK_{0IJ} \right) \neq 0$, because otherwise a linear combination of $v_I$ would have vanishing intersection with the limit direction and then it would lie in the $\ker \bK$. Moreover $\det \left( \cK_{\m\n a} M^a \right)$ is a moduli-dependent quantity and it is in general non-vanishing. The next step to compute the scaling of $\cK^{ab}$ is to compute the matrix of cofactors. From now on it will be enough to keep track of the scaling of the entries of the matrix, rather than the precise coefficients. At leading order we have
\be
\text{cof } \cK_{ab} \sim 
\begin{pmatrix}
    \phi^{-r} & \phi^{-r-1} & \phi^{-r-1} \\
    \phi^{-r-1} & \phi^{-r} & \phi^{-r}\\
    \phi^{-r-1} & \phi^{-r} & \phi^{-r+1}
\end{pmatrix} \, ,
\ee
where for some entries there might be some numerical cancellation that further suppress the scaling with $\phi$, but this would not spoil our argument. Finally, the inverse matrix is given by the cofactor matrix divided by the determinant and scales like
\be
\cK^{ab} \sim
\begin{pmatrix}
    \text{const} & \phi^{-1} & \phi^{-1} \\
    \phi^{-1} & \text{const} & \text{const} \\
    \phi^{-1} & \text{const} & \phi \\
\end{pmatrix}\, .
\ee
Now we are ready to argue that any divergent contribution to the moduli space curvature must come purely from the extended RFT sector $\{\bm{v}_I,\bm{v}_\m\}$. Indeed, one can check that any term with an index $0$ in the expression
\be     \label{R_divergent}
\cK^{ab} \cK^{cd} \cK^{fg} \cK_{ac[f} \cK_{b]dg} \, .
\ee
gives a contribution to the curvature that is at most constant. We show this explicitly for the case where the first index is $0$
\be
\cK^{0b} \cK^{cd} \cK^{fg} \cK_{0c[f} \cK_{b]dg} = \cK^{0b} \cK^{0d} \cK^{fg} \cK_{00[f} \cK_{b]dg} + \cK^{0b} \cK^{Id} \cK^{fg} \cK_{0I[f} \cK_{b]dg}\, ,
\ee
where we used the fact that $\cK_{0\m a}=0, \, \forall a$. In the first term, if $b \neq 0$ or $d \neq 0$ we already have at least a factor of $\phi^{-1}$, which cannot be compensated by any component $\cK^{fg}$ to get a divergence. The same happens in the second term, unless $b=0$ and $d \neq 0$. So we are left with the possible divergent contributions
\be
\cK^{00} \cK^{00} \cK^{fg} \cK_{00[f} \cK_{0]0g} + \cK^{00} \cK^{IJ} \cK^{fg} \cK_{0I[f} \cK_{0]Jg} + \cK^{00} \cK^{I\m} \cK^{fg} \cK_{0I[f} \cK_{0]\m g}\, .
\ee
But then, due to the properties of $\cK_{abc}$ in this basis, we are forced to take $f,g\neq\m$ to get something non-vanishing and the whole expression is at most constant.

\subsubsection*{$w=2$ limits}

In this class of limits we pick the basis of divisors $\{\bm{v}_E, \bm{v}_0, \bm{v}_I, \bm{v}_\m\}$ defined by
\be     \label{basisw2}
v_E^a {\bf k}_a \neq 0 \, , \qquad \bm{v}_0 = \bm{e}_0\, , \qquad \bm{v}_I^a {\bf k}_a = 0\, , \text{but } \bm{v}_I \notin \ker \bK \oplus \langle \bm{e}_0 \rangle \, , \qquad \bm{v}_\m \in \ker \bK \, .
\ee
In the general case we will have $\cK_{0EE},\cK_{0EI} \neq 0$, but if that is the case one can perform the shifts
\be
\bm{v}_E \to \bm{v}_E - \frac{\cK_{0EE}}{2\cK_{00E}} \bm{v}_0\, , \qquad \bm{v}_I \to \bm{v}_I - \frac{\cK_{0EI}}{\cK_{00E}} \bm{v}_0 \, ,
\label{apvsw2}
\ee
which are compatible with the definitions of our basis elements and ensure that $\cK_{0EE}=\cK_{0EI}=0$. Let us notice that these conditions are ensured if one picks the same basis as in equation (2.17) of \cite{Grimm:2013oga}. In their language, $\bm{v}_E$ would be identified with the shifted zero-section, the $\bm{v}_I$'s can be either exceptional divisors or images under the Shioda map of extra rational sections and $\bm{v}_0$ would be a combination of vertical divisors. In the basis described above we have
\be
\cK_{ab} \simeq
\begin{pmatrix}
    \cK_{EEa} M^a & \cK_{00E} M^0 & \cK_{EIa} M^a & \cK_{E\m a} M^a \\
    \cK_{00E} M^0 & \cK_{00E} M^E & \cK_{0I K} M^K & 0 \\
    \cK_{EIa} M^a & \cK_{0I K} M^K & \cK_{0IJ} M^0 & \cK_{I\m a} M^a \\
    \cK_{E\m a} M^a & 0 & \cK_{I\m a} M^a & \cK_{\m\n a} M^a
\end{pmatrix}
\sim
\begin{pmatrix}
    \phi^{-\frac{2}{3}} & \phi^\frac{1}{3} & \phi^{-\frac{2}{3}} & \phi^{-\frac{2}{3}} \\
    \phi^\frac{1}{3} & \phi^{-\frac{2}{3}} & \phi^{-\frac{2}{3}} & 0 \\
    \phi^{-\frac{2}{3}} & \phi^{-\frac{2}{3}} & \phi^\frac{1}{3} & \phi^{-\frac{2}{3}} \\
    \phi^{-\frac{2}{3}} & 0 & \phi^{-\frac{2}{3}} & \phi^{-\frac{2}{3}}
\end{pmatrix} \, ,
\ee
\be
\cK_a \simeq \left( \cK_{00E} (M^0)^2 \, , \, \cK_{00E} M^0 M^E \, , \,  \cK_{0IJ} M^0 M^J \, , \, \cK_{\m ab} M^a M^b \right) \sim \left( \phi^\frac{2}{3} \, , \phi^{-\frac{1}{3}} \, , \phi^{-\frac{1}{3}} \, , \phi^{-\frac{4}{3}} \right)\, ,
\ee
where the sums over $a$ do not include the direction of the limit $M^0$ and we have used that $M^0 \sim \phi^\frac{1}{3}$ and $M^a \sim \phi^{-\frac{2}{3}}, \, \forall a \neq 0$. The gauge kinetic matrix then scales like
\be
\cI_{ab} \sim \begin{pmatrix}
   \phi^\frac{4}{3} & \phi^\frac{1}{3} & \phi^\frac{1}{3} & \phi^{-\frac{2}{3}} \\
   \phi^\frac{1}{3} & \phi^{-\frac{2}{3}} & \phi^{-\frac{2}{3}} & \phi^{-\frac{5}{3}} \\
   \phi^\frac{1}{3} & \phi^{-\frac{2}{3}} & \phi^\frac{1}{3} & \phi^{-\frac{2}{3}} \\
   \phi^{-\frac{2}{3}} & \phi^{-\frac{5}{3}} & \phi^{-\frac{2}{3}} & \phi^{-\frac{2}{3}}
\end{pmatrix} \, ,
\ee
and the charge-to-tension ratios of BPS strings
\be
\g^2_E, \g^2_0 \sim \text{const} \, , \qquad \g^2_I \sim \phi \, , \qquad \g^2_\m \sim \phi^2 \, .
\ee
In this case, the behaviour of the ratio for a string charged under more than one of the sector is more complicated. In most cases the dominant contribution is the least diverging $\g$ from the ones above, except for the fact that the sector $I$ dominates over the limit direction. So we can have a string charged under both the sectors $\bm{v}_I$ and $\bm{v}_0$ (and possibly $\bm{v}_\m$) and still have $\g^2 \sim \phi$. From the previous analysis one can conclude that the direction $\bm{v}_0$ corresponds to the graviphoton multiplet and $\bm{v}_E$ belongs to the gravitational sector, while $\bm{v}_I$ and $\bm{v}_\m$ spanning the extended RFT field directions, with the core RFT given by the kernel sector $\bm{v}_\m$.

Again, we argue that any divergence in the curvature comes from the extended RFT sector. In order to do that we start by computing the determinant of $\cK_{ab}$
\be
\det \cK_{ab} \simeq - \left( \cK_{00E}\right)^2 \det \left( \cK_{0IJ} \right) \det \left( \cK_{\m\n a} M^a \right) \left(M^0\right)^{n_I+2} \sim \phi^{\frac{1}{3}(2+n_I-2r)} \, .
\ee
As before, $\det \left( \cK_{0IJ} \right) \neq 0$, because otherwise a linear combination of $v_I$ would be in the kernel and $\det \left( \cK_{\m\n a} M^a \right)$ is moduli dependent, so it is in general non-vanishing. The matrix of cofactors scales like
\be
\text{cof } \cK_{ab} \sim 
\begin{pmatrix}
    \phi^{\frac{1}{3}(n_I-2r-2)} & \phi^{\frac{1}{3}(n_I-2r+1)} & \phi^{\frac{1}{3}(n_I-2r-2)} & \phi^{\frac{1}{3}(n_I-2r-2)} \\
    \phi^{\frac{1}{3}(n_I-2r+1)} & \phi^{\frac{1}{3}(n_I-2r-2)} & \phi^{\frac{1}{3}(n_I-2r-2)} & \phi^{\frac{1}{3}(n_I-2r+1)}\\
    \phi^{\frac{1}{3}(n_I-2r-2)} & \phi^{\frac{1}{3}(n_I-2r-2)} & \phi^{\frac{1}{3}(n_I-2r+1)} & \phi^{\frac{1}{3}(n_I-2r+1)}\\
    \phi^{\frac{1}{3}(n_I-2r-2)} & \phi^{\frac{1}{3}(n_I-2r+1)} & \phi^{\frac{1}{3}(n_I-2r+1)} & \phi^{\frac{1}{3}(n_I-2r+4)}
\end{pmatrix} \, ,
\ee
and then the inverse matrix
\be
\cK^{ab} \sim
\begin{pmatrix}
    \phi^{-\frac{4}{3}} & \phi^{-\frac{1}{3}} & \phi^{-\frac{4}{3}} & \phi^{-\frac{4}{3}} \\
    \phi^{-\frac{1}{3}} & \phi^{-\frac{4}{3}} & \phi^{-\frac{4}{3}} & \phi^{-\frac{1}{3}} \\
    \phi^{-\frac{4}{3}} & \phi^{-\frac{4}{3}} & \phi^{-\frac{1}{3}} & \phi^{-\frac{1}{3}} \\
    \phi^{-\frac{4}{3}} & \phi^{-\frac{1}{3}} & \phi^{-\frac{1}{3}} & \phi^\frac{2}{3}
\end{pmatrix}\, .
\ee
Now we can show that any divergence in the curvature must come from the sector $\{\bm{v}_I,\bm{v}_\m\}$. First of all, notice that any time there is an index $0$ in \eqref{R_divergent} there must be two more indices that do not belong to the kernel, in order to have $\cK_{0ab} \neq 0$. But this implies at least a suppression factor of $\phi^{-\frac{2}{3}}$, which makes the curvature contribution at most constant. Then no index $0$ can appear in any divergent contribution. Let us now turn to the index $E$. If it appears in $\cK^{EE}, \cK^{EI}$ or $\cK^{E\m}$ it already gives a factor of $\phi^{-\frac{4}{3}}$, which can at most be compensated by two $\cK^{\m\n}$ components to give a constant piece, while if it appears in $\cK^{0E}$ we also have an index $0$ and the previous argument applies. We then conclude that only the extended RFT sector $\{\bm{v}_I,\bm{v}_\m\}$ can give rise to divergences.

\subsubsection*{$w=1$ limits}

In this class of limits we pick the basis of divisors $\{\bm{v}_0, \bm{v}_I, \bm{v}_\m\}$ defined by
\be
\bm{v}_0 = \bm{e}_0\, , \qquad \bm{v}_I \notin \ker \bK\, , \qquad \bm{v}_\m \in \ker \bK \, , \text{but } \bm{v}_\m \neq \bm{e}_0 \, .
\label{apvsw1}
\ee
At leading order the matrix $\cK_{ab}$ takes the form
\be
\cK_{ab} \simeq
\begin{pmatrix}
    0 & \cK_{0IJ} M^J & 0 \\
    \cK_{0IJ} M^J & \cK_{0IJ} M^0 & \cK_{I\m a} M^a \\
    0 & \cK_{I\m a} M^a & \cK_{\m\n a} M^a
\end{pmatrix}
\sim
\begin{pmatrix}
    0 & \phi^{-\frac{1}{3}} & 0 \\
    \phi^{-\frac{1}{3}} & \phi^\frac{2}{3} & \phi^{-\frac{1}{3}} \\
    0 & \phi^{-\frac{1}{3}} & \phi^{-\frac{1}{3}}
\end{pmatrix} \, ,
\ee
\be
\cK_a \simeq \left( \cK_{0IJ} M^I M^J \, , \,  \cK_{0IJ} M^0 M^J \, , \, \cK_{\m ab} M^a M^b \right) \sim \left( \phi^{-\frac{2}{3}} \, , \phi^\frac{1}{3} \, , \phi^{-\frac{2}{3}} \right)\, ,
\ee
where the sums over $a$ do not include the direction of the limit $M^0$ and we have used that $M^0 \sim \phi^\frac{2}{3}$ and $M^a \sim \phi^{-\frac{1}{3}}, \, \forall a \neq 0$. The gauge kinetic matrix then scales like
\be
\cI_{ab} \sim \begin{pmatrix}
    \phi^{-\frac{4}{3}} & \phi^{-\frac{4}{3}} & \phi^{-\frac{4}{3}} \\
    \phi^{-\frac{4}{3}} & \phi^\frac{2}{3} & \phi^{-\frac{1}{3}} \\
    \phi^{-\frac{4}{3}} & \phi^{-\frac{1}{3}} & \phi^{-\frac{1}{3}}
\end{pmatrix} \, ,
\ee
and the charge-to-tension ratios of BPS strings
\be
\g^2_0 \sim \text{const} \, , \qquad \g^2_I \sim \text{const} \, , \qquad \g^2_\m \sim \phi \, .
\ee
In this class of limits, a string that is charged under both the kernel sector and the limit direction still has a divergent $\g$, while as soon as it has a non-vanishing charge in the $I$ sector, $\g$ becomes finite. From these expressions, we can see that the limit direction $\bm{v}_0$ and the directions $\bm{v}_I$ are part of the gravitational sector, while the extended and core RFT coincide and contain just the sector $\bm{v}_\m$. We now proceed to show that any curvature divergence is sourced by the core RFT. We start by computing the determinant
\be
\det \cK_{ab} \simeq - \left( \cK_{0KL} M^K M^L \right) \det \left( \cK_{0IJ} \right) \det \left( \cK_{\m\n a} M^a \right) \left( M^0 \right)^{n_I-1} \sim \phi^{-\frac{1}{3}(4+r-2n_I)}\, ,
\ee
where all the factors are different from zero for similar arguments as before and the matrix of cofactors
\be
\text{cof } \cK_{ab} \sim 
\begin{pmatrix}
    \phi^{-\frac{1}{3}(r-2n_I)} & \phi^{-1-\frac{1}{3}(r-2n_I)} & \phi^{-1-\frac{1}{3}(r-2n_I)} \\
    \phi^{-1-\frac{1}{3}(r-2n_I)} & \phi^{-2-\frac{1}{3}(r-2n_I)} & \phi^{-2-\frac{1}{3}(r-2n_I)} \\
    \phi^{-1-\frac{1}{3}(r-2n_I)} & \phi^{-2-\frac{1}{3}(r-2n_I)} & \phi^{-1-\frac{1}{3}(r-2n_I)}
\end{pmatrix} \, .
\ee
Finally the inverse matrix scales like
\be
\cK^{ab} \sim
\begin{pmatrix}
    \phi^\frac{4}{3} & \phi^\frac{1}{3} & \phi^\frac{1}{3} \\
    \phi^\frac{1}{3} & \phi^{-\frac{2}{3}} & \phi^{-\frac{2}{3}} \\
    \phi^\frac{1}{3} & \phi^{-\frac{2}{3}} & \phi^\frac{1}{3} \\
\end{pmatrix}\, .
\ee
Similarly to what we have done before, one can check explicitly that any term containing an index that does not belong to the sector $\{\bm{v}_\m\}$ leads to at most a constant contribution to the curvature.

\section{The moduli space metric in rigid limits}
\label{ap:metric}

In this section we show that the RFT metric coincides with its gauge kinetic matrix at leading order
\be
g_{m n} \simeq \frac{1}{2} \cI_{mn} \simeq - \frac{1}{2} \cK_{mn}\, .
\ee
Let us start from the definition
\be     \label{gij}
g_{ij} = \frac{1}{2} \cI_{ab} \frac{\pa M^a}{\pa \psi^i} \frac{\pa M^b}{\pa \psi^j} \, ,
\ee
where we will restrict to the rigid sector $g_{mn}$ for which $\cI_{mn} \simeq -\cK_{mn}$. Moreover, as explained in section \ref{s:rigid}, in a rigid limit the direction of the limit is always decoupled and never belongs to the RFT sector. So we can pick as coordinates
\be
\psi^i = M^i \, , \qquad i \neq e \, ,
\ee
and we use the constraint $\cV_X = 1$ to express $M^e$ as a function of $\psi^i$. Then we have
\be
\frac{\pa M^a}{\pa \psi^i} = \left( \frac{\pa M^e}{\pa M^i}, \del_i^j \right)\, ,
\ee
where we can compute $\frac{\pa M^e}{\pa M^i}$ by using the constant volume constraint
\be
\frac{\d \cV_X}{\d M^i} = \frac{\pa \cV_X}{\pa M^i} + \frac{\pa \cV_X}{\pa M^e} \frac{\pa M^e}{\pa M^i} = 0 \qquad \implies \qquad \frac{\pa M^e}{\pa M^i} = -\frac{\cK_i}{\cK_e} \, .
\ee
Substituting everything into \eqref{gij} and restricting to the rigid sector we get
\be     \label{gmn}
g_{mn} = \frac{1}{2} \cI_{mn} - \frac{1}{2} \left( \cI_{en} \frac{\cK_m}{\cK_e} + \cI_{em} \frac{\cK_n}{\cK_e} \right) + \frac{1}{2} \cI_{ee} \frac{\cK_m \cK_n}{\cK_e^2} \, .
\ee
Now we want to show that the second and third terms in this expression are always subleading with respect to the first one, which at leading order reduces to $\cK_{mn}$. We divide the discussion into three sections, according to the scaling index $w$ of the limit we consider. For each class of limit, we use the adapted basis defined in Appendix \ref{ap:asymptotic}.\\

\noindent {\bf $w=3$ limits}

\noindent Here the extended rigid sector contains the directions $\{M^m\} = \{M^I, M^\m\}$ and we have
\be
\begin{split}
\cK_{\m\n}, \cK_{\m I} \sim \phi^{-1} \, , \qquad &\cK_{IJ} \sim \text{const} \, , \qquad \cK_e \sim \text{const} \, , \qquad \cK_I \sim \phi^{-1} \, , \qquad \cK_\m \sim \phi^{-2} \, , \\
&\cI_{e\m} \sim \phi^{-2} \, , \qquad \cI_{eI} \sim \phi^{-1} \, , \qquad \cI_{ee} \sim \text{const} \, .
\end{split}
\ee
We see that $\cK_{mn}$ scales either as $\phi^{-1}$ or as a constant, while all the other terms in \eqref{gmn} are at least suppressed by a factor of $\phi^{-2}$. Thus we conclude that
\be
g_{mn} \simeq -\frac{1}{2} \cK_{mn} \, , \qquad m,n \in \{I,\m\} \, .
\ee

\noindent {\bf $w=2$ limits}

\noindent As in the previous case, the field directions that belong to the extended RFT are $\{M^m\} = \{M^I, M^\m\}$ and we have the scalings
\be
\begin{split}
\cK_{\m\n}, \cK_{\m I} \sim \phi^{-\frac{2}{3}} \, , \qquad &\cK_{IJ} \sim \phi^{\frac{1}{3}} \, , \qquad \cK_e, \cK_I \sim \phi^{-\frac{1}{3}} \, , \qquad \cK_\m \sim \phi^{-\frac{4}{3}} \, , \\
&\cI_{e\m} \sim \phi^{-\frac{5}{3}} \, , \qquad \cI_{eI}, \cI_{ee} \lesssim \phi^{-\frac{2}{3}} \, .
\end{split}
\ee
We see that $\cK_{mn}$ scales either as $\phi^{\frac{1}{3}}$ or as $\phi^{-\frac{2}{3}}$, while all the other terms in \eqref{gmn} are at least suppressed by a factor of $\phi^{-\frac{2}{3}}$. However, the only components for which the latter can compete with the first term are $m=I,n=J$, where the corrections scale as $\phi^{-\frac{2}{3}}$, but for these the first term scales like $\cK_{IJ} \sim \phi^\frac{1}{3}$. Thus we conclude that
\be
g_{mn} \simeq -\frac{1}{2} \cK_{mn} \, , \qquad m,n \in \{I,\m\} \, .
\ee

\noindent {\bf $w=1$ limits}

\noindent In the last class of limit the RFT only contains the kernel directions $\{M^\m\}$ and we have
\be
\cK_{\m\n} \sim \phi^{-\frac{1}{3}} \, , \qquad \cK_e, \cK_\m \sim \phi^{-\frac{2}{3}} \, , \qquad \cI_{e \m}, \cI_{ee} \sim \phi^{-\frac{4}{3}} \, .
\ee
We see that all the other terms in \eqref{gmn} are at least suppressed by a factor of $\phi^{-\frac{4}{3}}$, which makes them subleading with respect to $\cK_{\m\n}$. Thus we conclude that
\be
g_{\m\n} \simeq -\frac{1}{2} \cK_{\m\n}\, .
\ee



\bibliographystyle{JHEP2015}
\bibliography{papers}

\end{document}